\documentclass[10pt]{article}
\usepackage[acronym]{glossaries}
\usepackage{physics}
\usepackage{siunitx}
\usepackage[version=4]{mhchem}
\usepackage[numbers,sort&compress]{natbib}
\usepackage{authblk}
\usepackage[margin=1in]{geometry}
\usepackage{xcolor}
\usepackage{graphicx}
\usepackage{hyperref}

\newcommand{\SEE}[1]{\textit{cf.}~{#1}}
\newcommand{\VEC}[1]{\mathbf{#1}}
\newcommand{\UNITVEC}[1]{\mathbf{\hat{#1}}}
\newcommand{\FIGTEXTDUMMY}[1]{Fig.~{#1}}
\newcommand{\FIGTEXTTWODUMMY}[2]{Figs.~{#1}~and~{#2}}
\newcommand{\FIGTEXT}[1]{\FIGTEXTDUMMY{\ref{fig:#1}}}
\newcommand{\FIGTEXTTWO}[2]{\FIGTEXTTWODUMMY{\ref{fig:#1}}{\ref{fig:#2}}}
\newcommand{\FIGTEXTSTART}[1]{Figure~\ref{fig:#1}}
\newcommand{\EQTEXT}[1]{Eq.~\ref{eq:#1}}
\newcommand{\EQTEXTTWO}[2]{Eqs.~\ref{eq:#1}~and~\ref{eq:#2}}
\newcommand{\EQTEXTSTART}[1]{Equation~\ref{eq:#1}}
\newcommand{\EQTEXTTWOSTART}[2]{Equations~\ref{eq:#1}~and~\ref{eq:#2}}
\newcommand{\TABTEXT}[1]{Table~\ref{tab:#1}}
\newcommand{\APPTEXT}[1]{Appendix~{\ref{sec:app:#1}}}
\newcommand{\DIM}[1]{{#1}}
\newcommand{\DIMCONST}[1]{\DIM{#1}}
\newcommand{\NONDIM}[1]{{{#1}^*}}
\newcommand{\NONDIMCONST}[1]{\NONDIM{#1}}
\newcommand{\PARTIAL}[2]{\frac{\partial #1}{\partial #2}}
\newcommand{\ORDER}[1]{\mathcal{O}\left({#1}\right)}

\DeclareSIUnit{\Molar}{\textsc{m}}
\DeclareSIUnit{\fps}{fps}

\definecolor{rmdarkblue}{HTML}{1864ab}
\hypersetup{colorlinks, citecolor=rmdarkblue, linkcolor=rmdarkblue, urlcolor=rmdarkblue}

\setacronymstyle{long-short}
\newacronym{cps}{cPS}{carboxyl-functionalized polystyrene}
\newacronym{aps}{aPS}{amine-functionalized polystyrene}
\newacronym{ps}{PS}{polystyrene}
\newacronym{pdms}{PDMS}{polydimethylsiloxane}
\newacronym{piv}{PIV}{particle image velocimetry}

\begin{document}

\title{Convection rolls and three-dimensional particle dynamics in merging solute streams}

\author[1]{Robben E. Migacz}
\author[1,a]{Guillaume Durey}
\author[1,b]{Jesse T. Ault}

\affil[1]{Center for Fluid Mechanics, School of Engineering, Brown University, Providence, Rhode Island 02912, USA}
\affil[a]{Current affiliation: European Organization for Nuclear Research (CERN), Esplanade des Particules 1, 1217 Meyrin, Switzerland}
\affil[b]{\href{mailto:jesse_ault@brown.edu}{jesse\_ault@brown.edu}}

\maketitle

\begin{abstract}
Microparticles migrate in response to gradients in solute concentration through diffusiophoresis and diffusioosmosis. Merging streams of fluid with distinct solute concentrations is a common strategy for producing a steady concentration gradient with continuous flow in microfluidic devices; the solute concentration gradient and consequent diffusiophoresis are primarily normal to the background flow. This is particularly useful in separation and filtration processes, as it results in regions of particle accrual and depletion in continuous flows. Such systems have been examined in several classic papers on diffusiophoresis, with a focus on the particle dynamics far from boundaries. We show, through experiments, simulations, and theory, that diffusioosmotic flow along certain boundaries can result in significant changes in particle dynamics and particle focusing in near-wall regions. The nonzero velocity at charged surfaces draws solute and particles along the boundary until the flow ultimately recirculates. These ``convection rolls,'' which result in the spanwise migration of polystyrene particles close to boundaries, are apparent near a glass surface but vanish when the surface is coated with gold. The three-dimensional nature of the dynamics could have implications for the design of microfluidic devices: Channel materials can be selected to enhance or suppress near-wall flows. Additionally, we demonstrate the importance of considering solute concentration-dependent models for diffusiophoretic and diffusioosmotic mobility in capturing the dynamics of particles, particularly in regions of low solute concentration.
\end{abstract}

\section{Introduction}
\begin{figure}
    \centering
    \includegraphics{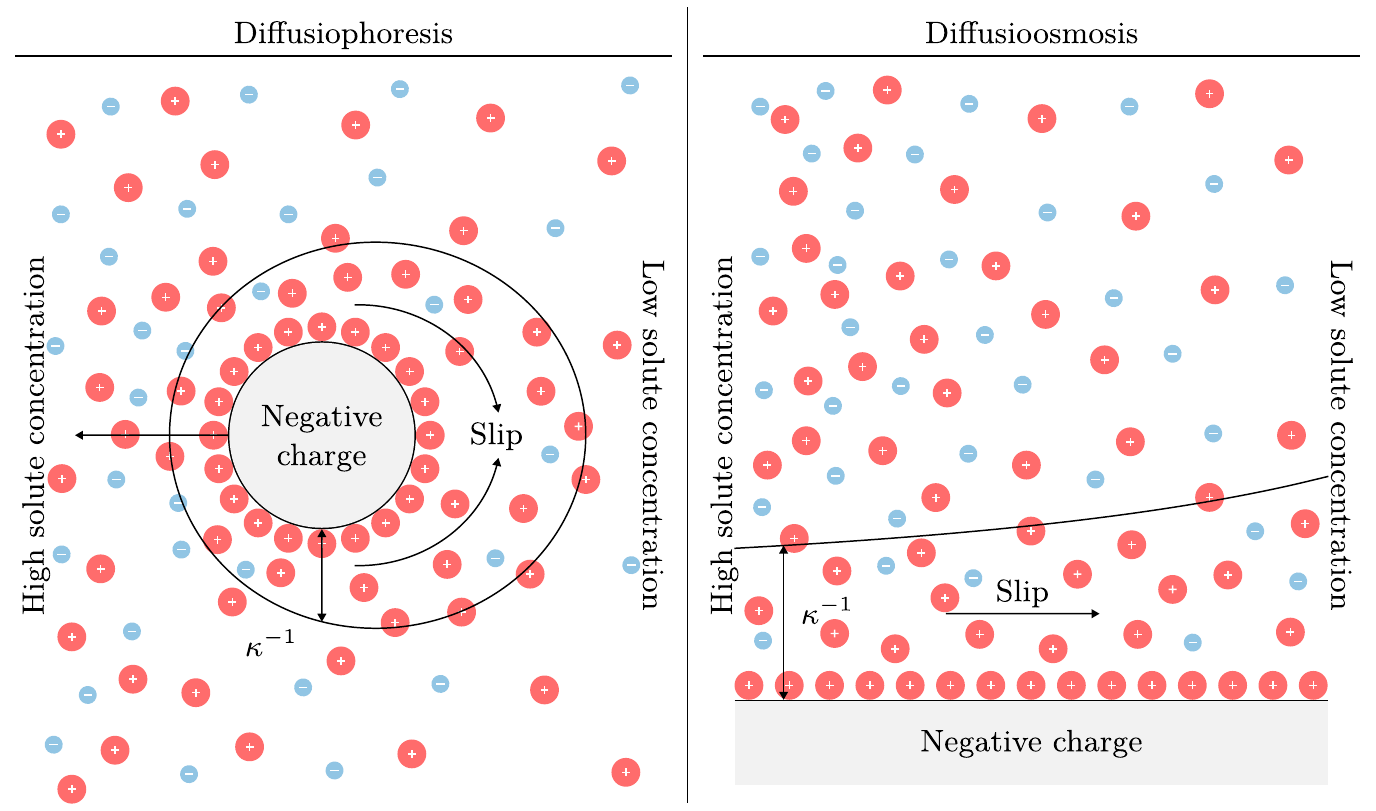}
    \caption{Mechanisms of diffusiophoresis and diffusioosmosis. Diffusiophoresis is the motion of a particle in response to a solute concentration gradient. Diffusioosmosis, a closely related phenomenon, occurs along a surface. In both cases, there is a slip flow along the surface that results from an osmotic pressure gradient and a spontaneous electric field that develops to maintain electroneutrality. The direction of motion is dependent on the particle or surface material and the species present in the flow. The species and the Debye length $\DIM{\kappa}^{-1}$ are not shown to scale.}\label{fig:diffusiophoresis-diffusioosmosis}
\end{figure}

Diffusiophoresis, the spontaneous migration of particles in response to a solute concentration gradient, is commonly observed or applied in systems with particles on the order of tens or hundreds of nanometers \cite{Deseigne2014,Abecassis2008,Seo2020,Lechnick1984,Alessio2022,Shin2017a,Shin2015,Shin2018,Palacci2010,Palacci2012,Ebel1988,Shi2016,Shimokusu2019,Singh2020} to several micrometers \cite{Shin2017,Seo2020,Timmerhuis2022,Shin2015,Shim2021,Kar2015,Shim2020,Banerjee2016,Florea2014,Alessio2021,McDermott2012,Shim2022a,Shin2017b,Wilson2020,Shimokusu2019,Battat2019,Yang2023} in diameter. Diffusioosmosis, a closely related phenomenon, results in the spontaneous development of a slip velocity along charged surfaces in the presence of a solute concentration gradient \cite{Shim2022}. The mechanism of diffusiophoresis and diffusioosmosis is shown in \FIGTEXT{diffusiophoresis-diffusioosmosis}; particle migration or flow along a surface is a result of both an osmotic pressure gradient and an electric field that develops to maintain neutrality when charged species would otherwise diffuse at different rates if they were neutral \cite{Shim2022}. First described in the mid-twentieth century \cite{Derjaguin1947,Derjaguin1961} with refined models and experimental validation in subsequent decades \cite{Anderson1982,Prieve1984,Lechnick1984,Staffeld1989}, diffusiophoresis and diffusioosmosis have received considerable attention in recent years. This increase in interest coincides largely with the growth of the field of microfluidics: Diffusiophoresis and diffusioosmosis allow for the controlled motion of species in a flow through the manipulation of solute concentration gradients rather than the application of an external field. This makes the phenomena highly relevant to various problems in microfluidic contexts, such as enhanced transport relative to diffusion in dead-end pores \cite{Shin2017,Alessio2022,Shin2017a,Shin2015,Kar2015,Shin2018,Alessio2021,Shim2022a,Singh2020,Battat2019,Tan2021,Akdeniz2023,Singh2022} or the separation of species or development of an exclusion zone \cite{Abecassis2008,Seo2020,Timmerhuis2022,Shin2020,Shim2021,Banerjee2016,Florea2014,Elton2020,Shi2016,Shin2017b,Shimokusu2019,Visan2019}.

Geometries with merging streams of fluid of distinct solute concentrations (co-flowing solutions) are commonly used to produce solute concentration gradients in experiments \cite{Abecassis2008,Shim2022,Velegol2016,Chakra2023,Timmerhuis2022,Shimokusu2019,Yang2023}. In such configurations, the concentration gradient is approximately perpendicular to the flow, which results in the transverse migration of particles through diffusiophoresis and, indirectly, diffusioosmosis. This feature is particularly useful in separation and filtration processes, as it yields regions of particle accrual and depletion in continuous flows. This has been examined in classic papers on diffusiophoresis, such as that by \citet{Abecassis2008}, which comments on particle focusing and spreading through diffusiophoresis in a ``$\psi$-channel,'' so named for its three-inlet geometry that resembles the letter. Similar channel designs with two or three merging inlet streams have been used in several studies \cite{Seo2020,Timmerhuis2022,Shin2020,Shin2017a,Chakra2023}. Such works tend to focus on particle migration at the center of the channel, far from boundaries, where the dominant mechanism for transport is diffusiophoresis.

This assumption of motion in reduced dimensions is common throughout the literature on diffusiophoresis and diffusioosmosis. \citet{Ault2017}, for instance, consider one-dimensional diffusiophoretic motion to describe particle migration in pores. They later model the motion of particles in quasi-one-dimensional pores of high aspect ratio, with numerical simulations and leading-order corrections to describe two-dimensional dynamics \cite{Ault2018}. Even where two- or three-dimensional particle motion is considered, such works often assume solute concentration gradients are one-dimensional \cite{Raj2023}. Several recent works, however, have considered solute and particle dynamics in two or three dimensions. \citet{Migacz2022} describe the two-dimensional dynamics of solutes and particles in a narrow channel, neglecting diffusioosmosis, and provide numerical results in three dimensions. A more recent work by \citet{TengDiffusioosmosis} follows a similar procedure to describe the effect of diffusioosmosis on solute diffusion, in two dimensions and in cylindrical coordinates, absent background flow and in a narrow channel. They show that the recirculating flow induced by the slip velocity at channel walls distorts the solute concentration profile and affects the rate of diffusion in a manner analogous to Taylor dispersion. These are complementary works that show how both diffusiophoresis (directly) and diffusioosmosis (indirectly) affect particle dynamics in two and three dimensions. This has recently been explored in dead-end pores: \citet{Alessio2022} and \citet{Akdeniz2023} consider the effect of diffusioosmotic flow on three-dimensional particle dynamics in a dead-end pore. Other recent works have similarly concluded that considerations of dynamics beyond one dimension are important. \citet{Chu2021} develop a macrotransport equation to approximate the dynamics of particles in a channel of uniform, circular cross-section by accounting for the effects of hydrodynamic dispersion, which would not be relevant in one dimension. The dynamics arising from both diffusiophoresis and diffusioosmosis in tandem have been explored in very recent work: \citet{Chakra2023} examine particle dynamics in a $\psi$-channel, accounting for both diffusiophoresis and diffusioosmosis, in work that was completed contemporaneously and independently; they comment on potential applications in particle separations and characterizations. A focusing effect that is similar to the one we demonstrate is also described in a recent work by \citet{Yang2023}, who demonstrate distinct particle dynamics in the presence of a surfactant gradient and complexing polymer. Such studies demonstrate the importance of considering the dynamics in multiple spatial dimensions and accounting for boundaries. Our work is, to our knowledge, the first to experimentally alter the surface charge of our channels to modulate the diffusioosmotic effects in a system of merging solute streams, as well as to systematically characterize these convection rolls via both theory and simulations.

Many simplifications beyond assumptions about solute or particle motion in reduced spatial dimensions are common to studies of diffusiophoresis and diffusioosmosis. One common simplification is an assumption that the diffusiophoretic or diffusioosmotic mobilities are constant. This has been revisited in recent years \cite{Migacz2022,LeeZetaPotential} because it overlooks potentially significant sources of variation; notable examples are the variation of zeta potential with conditions such as solute concentration, temperature, and pH \cite{Kirby2004}, and the role of size effects \cite{Shin2015}. A recent work by \citet{Akdeniz2023} demonstrates that concentration-dependent zeta potentials have a significant effect on particle dynamics in dead-end pore geometries where both diffusiophoresis and diffusioosmosis are considered. Other recent works, such as that by \citet{Shim2022a}, consider the effect of pH gradients on diffusiophoretic motion and demonstrate the importance of accounting for the local conditions when characterizing particle dynamics.

In this work, we focus on the three-dimensional dynamics of particles and their importance in through-flow systems, accounting for variable zeta potential, diffusiophoretic mobility, and diffusioosmotic mobility. To examine the dynamics of particles near boundaries, we study \gls{ps} particles in gradients of \ce{NaCl} produced by merging streams of distinct concentration; these are species common to numerous experimental studies of diffusiophoresis and diffusioosmosis \cite{Shin2017,Seo2020,Alessio2022,Shin2017a,Shin2015,Kar2015,Palacci2010,Palacci2012,Ebel1988,Florea2014,Shi2016,Alessio2021,Wilson2020,Battat2019,Shah2022,Paustian2015,Akdeniz2023,Yang2023}. We consider the dynamics of particles near glass and gold surfaces. Glass develops a surface charge in water through reactions at the surface \cite{Behrens2001}; gold is a noble metal and does not react with water. Therefore, we anticipate that diffusioosmotic transport is significant near glass but negligible near gold. In this context, we demonstrate that diffusioosmosis is an important consideration where solute gradients are present in microfluidic devices fabricated from glass or other materials of nonzero surface charge, which are common throughout the literature on diffusiophoresis and related phenomena \cite{Timmerhuis2022,Alessio2022,Shin2017a,Shin2015,Shim2021,Kar2015,Rasmussen2020,Shim2020,Palacci2010,Palacci2012,Banerjee2016,Ebel1988,Florea2014,Shi2016,Alessio2021,McDermott2012,Shim2022a,Shin2017b,Doan2020,Wilson2020,Shimokusu2019,Battat2019,McDonald2000,Sia2003,Peter2022,Akdeniz2023,Yang2023}, and show that the particle dynamics near glass and gold surfaces are distinct. Near the glass surface, fluid is drawn toward the center of the channel, where it is advected away from the boundary, forming swirling regions we call ``convection rolls.'' We demonstrate, through experiments and simulations, the relevance of three-dimensional dynamics and solute concentration-dependent models for diffusiophoretic and diffusioosmotic mobility. Additionally, we provide first-order velocity and solute concentration profiles, along with sample particle trajectories, in \APPTEXT{analytical}. The three-dimensional particle dynamics are particularly relevant to mixing processes; depending on the context, suppression or enhancement of the convection rolls phenomenon may be desirable to researchers and engineers working with microfluidic devices.

\section{Methods}
\begin{figure}
    \centering
    \includegraphics{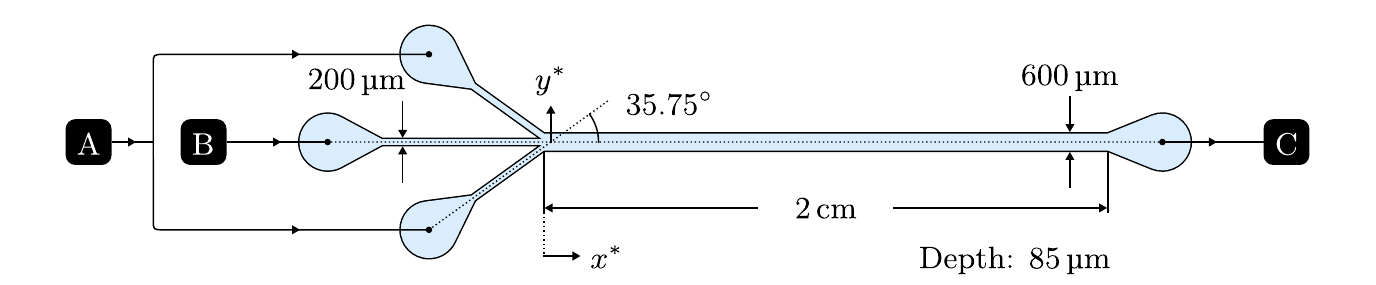}
    \caption{Geometry of the channel, viewed from the top. Each of the three inlets is $\SI{200}{\micro\meter}$ wide; the channel is $\SI{600}{\micro\meter}$ wide after the inlet streams merge and approximately $\SI{85}{\micro\meter}$ deep (out-of-plane) throughout. The outlet region is $\SI{2}{\centi\meter}$ from the point at which the streams merge at $\NONDIM{x}=0$. We introduce a $\SI{20}{\milli\Molar}$ solution of \ce{NaCl} to (A) at $\SI{1}{\micro\liter\per\minute}$ and deionized water with $\SI{0.01}{\percent}$ particles by mass to (B) at $\SI{1}{\micro\liter\per\minute}$. The flow into (A) is split between two inlets. We add $\SI{0.1}{\percent}$ TWEEN{\textregistered} 80 by mass to inhibit particle adhesion to the walls of the channel. The flow exits the channel at (C). The large circular regions at each inlet and at the outlet allow us to connect tubing to the device.}\label{fig:channel}
\end{figure}

We study the dynamics of fluid, solutes, and particles in $\psi$-channels, depicted in \FIGTEXT{channel}, where each of the three inlets is $\SI{200}{\micro\meter}$ wide and leads to a primary channel $\SI{2}{\centi\meter}$ in length and $\SI{600}{\micro\meter}$ in width. The height of the channel is approximately $\SI{85}{\micro\meter}$ throughout. For convenience, we introduce the notation $\DIM{x} = \left( \SI{2}{\centi\meter} \right) \NONDIM{x}$, $\DIM{y} = \left( \SI{600}{\micro\meter} \right) \NONDIM{y}$, and $\DIM{z} = \left( \SI{85}{\micro\meter} \right) \NONDIM{z}$, with $\NONDIM{y} = \NONDIM{z} = 0$ in the center of the channel and $\NONDIM{x} = 0$ where the inlets merge, such that $\NONDIM{x} \in \left[ 0, 1 \right]$, $\NONDIM{y} \in \left[ -\frac{1}{2}, \frac{1}{2} \right]$, and $\NONDIM{z} \in \left[ -\frac{1}{2}, \frac{1}{2} \right]$.

\subsection{Experiments}
We fabricate channels with conventional soft lithography techniques. The microchannel is composed of \gls{pdms} on all but the bottom surface, which is either glass or a $\SI{70}{\nano\meter}$-thick layer of gold over a $\SI{30}{\nano\meter}$-thick layer of chromium on glass. We use a plasma cleaner to bond the \gls{pdms} to the bottom surface of the device. We use \gls{ps}, \gls{cps}, and \gls{aps} particles in experiments; the specific particles used are presented in \APPTEXT{experimental-materials}. We introduce $\SI{20}{\milli\Molar}$ \ce{NaCl} to the distal inlets, shown in \FIGTEXT{channel}(A), and $\SI{0.01}{\percent}$ particles by mass to the center inlet, shown in \FIGTEXT{channel}(B). The solutions are introduced to (A) and (B), each at $\SI{1}{\micro\liter\per\minute}$. The flow into (A) is split between two inlets. We also include TWEEN{\textregistered} 80 surfactant at $\SI{0.1}{\percent}$ by mass to prevent the adhesion of particles to channel walls.

We capture experimental images with a Leica DMi8 microscope with a $20\times$ objective and $4\times$ binning. In experiments with a glass bottom surface, we use an exposure time of $\SI{100}{\milli\second}$ and a framerate of $\SI{10}{\fps}$; in experiments with a gold bottom surface, we instead use an exposure of $\SI{40}{\milli\second}$ with a framerate of $\SI{25}{\fps}$ to account for changes in lighting. The flow in the channel is driven by a Harvard Apparatus PHD ULTRA{\texttrademark} syringe pump with two Hamilton{\texttrademark} $\SI{100}{\micro\liter}$ syringes. The flow into (A) is split across two inlets. We perform \gls{piv} with PIVlab 2.56 \cite{Thielicke2014,Thielicke2021} to determine the transverse velocity of particles near the bottom surface of the channel. The electrophoretic mobility of \gls{cps} particles was measured by Viet Sang Doan, University at Buffalo, with an Anton Paar Litesizer{\texttrademark} 500.

\subsection{Simulations}
We model the structures observed in the particle concentration fields by performing finite volume simulations of the system. Assuming density is constant and inertial effects are negligible, the steady-state flow is governed by
\begin{align}
\DIMCONST{\VEC{\nabla}} \cdot \DIM{\VEC{u}} &= 0\quad{\rm and} \label{eq:continuity}\\
\DIMCONST{\VEC{\nabla}} \DIM{p} - \DIMCONST{\mu} \DIMCONST{\VEC{\nabla}}^2 \DIM{\VEC{u}} &= \VEC{0}, \label{eq:momentum}
\end{align}
where $\DIM{\VEC{u}}$ is the fluid velocity, $\DIM{\rho}$ is its density and $\DIM{\mu}$ its viscosity, $\DIM{t}$ is time, and $\DIM{p}$ is pressure. We assume the Stokes number of the particles is low, such that they would act as tracer particles for the background flow in the absence of diffusiophoresis. We also assume that the presence of the particles does not affect the background flow. Assuming diffusivity is constant and employing \EQTEXT{continuity}, the solute and particle dynamics are governed by
\begin{align}
\PARTIAL{\DIM{c}}{\DIM{t}} - \DIMCONST{{\mathcal{D}}_{c}} \DIMCONST{\VEC{\nabla}}^2 \DIM{c} + \DIM{\VEC{u}} \cdot \DIMCONST{\VEC{\nabla}} \DIM{c} &= 0\quad{\rm and} \label{eq:solute-dynamics}\\
\PARTIAL{\DIM{n}}{\DIM{t}} - \DIMCONST{{\mathcal{D}}_{n}} \DIMCONST{\VEC{\nabla}}^2 \DIM{n} + \DIMCONST{\VEC{\nabla}} \cdot \left[ \left( \DIM{\VEC{u}} + \DIM{{\mathcal{M}}} \DIMCONST{\VEC{\nabla}} \ln \NONDIM{c} \right) \DIM{n} \right] &= 0, \label{eq:particle-dynamics}
\end{align}
where $\DIM{c}$ is the solute concentration; $\DIMCONST{{\mathcal{D}}_{c}}$ is the ambipolar diffusivity of the solute \cite{Shin2015},
\begin{equation}
\DIMCONST{{\mathcal{D}}_{c}} = \frac{2 \DIMCONST{{\mathcal{D}}_{+}} \DIMCONST{{\mathcal{D}}_{-}}}{\DIMCONST{{\mathcal{D}}_{+}} + \DIMCONST{{\mathcal{D}}_{-}}},
\end{equation}
with cationic and anionic diffusivities $\DIMCONST{{\mathcal{D}}_{+}}$ and $\DIMCONST{{\mathcal{D}}_{-}}$, respectively; $\DIM{n}$ is the particle concentration; $\DIMCONST{{\mathcal{D}}_{n}}$ is the particle diffusivity; and $\DIM{{\mathcal{M}}}$ is the diffusiophoretic mobility. We use the dimensionless solute concentration $\NONDIM{c} = \frac{\DIM{c}}{\SI{1}{\Molar}}$ for notational convenience, though the choice is ultimately immaterial because the scaling of $\DIM{c}$ in $\DIMCONST{\VEC{\nabla}} \ln \NONDIM{c} = \frac{\DIMCONST{\VEC{\nabla}} \NONDIM{c}}{\NONDIM{c}}$ will cancel. We calculate the particle diffusivity with the Stokes--Einstein relation \cite{Miller1924},
\begin{equation}
\DIMCONST{{\mathcal{D}}_{n}} = \frac{\DIMCONST{{k_{\text{B}}}} \DIMCONST{T}}{6 \pi \DIMCONST{\mu} \DIMCONST{a}}.
\end{equation}
Here, $\DIMCONST{{k_{\text{B}}}}$ is the Boltzmann constant, $\DIMCONST{T}$ is the absolute temperature, and $\DIMCONST{a}$ is the particle radius. \EQTEXTTWOSTART{solute-dynamics}{particle-dynamics} are advection--diffusion equations with an additional component of velocity for diffusiophoresis of the particles. At boundaries, we impose
\begin{equation}
\left. \DIM{\VEC{u}} \right|_{\rm b} = \begin{cases}
\DIM{\VEC{u}_{{{\text{do}}}}} = \DIM{{\mathcal{M}}_{{{\text{do}}}}} \DIMCONST{\VEC{\nabla}} \ln \NONDIM{c}, & {\text{if the surface is glass}}\\
\VEC{0}, & {\text{otherwise}},
\end{cases}
\quad{\rm with}\quad \left. \DIM{\VEC{u}} \right|_{\rm b} \cdot \UNITVEC{n} = 0, \label{eq:boundaries}
\end{equation}
where $\DIM{{\mathcal{M}}_{{{\text{do}}}}}$ is the diffusioosmotic mobility and $\UNITVEC{n}$ is a unit vector normal to the surface. This ensures the diffusioosmotic velocity is in the plane of the boundary and there is no fluid flow through the walls.

We implement a variable zeta potential for both the particles and glass surface. The zeta potential of \gls{cps} particles is shown as a function of solute concentration in \APPTEXT{zetamobilitymodels}; we have calculated the particle zeta potential from experimental measurements of electrophoretic mobility. We use the semi-analytical model of \citet{Ohshima1983}, which accounts for convective ion migration \cite{KirbyBook}, to relate experimentally obtained electrophoretic mobility measurements to the particle zeta potential. We then apply the model of \citet{Keh2000} to calculate the diffusiophoretic mobility with the modeled zeta potential. For the boundaries of the channel, we use a model where the wall zeta potential, $\DIM{\zeta}_{\rm b}$, is proportional to $\log \NONDIM{c}$, with various proportionality constants to cover a range of values given for silica in the literature. We provide further details about the models for zeta potential and mobility and compare with published values in \APPTEXT{zetamobilitymodels}.


We perform simulations to determine the steady-state solute and particle concentration fields. We use OpenFOAM \cite{Weller1998} to simulate the flow with components from simpleFoam to calculate the background flow described by \EQTEXTTWO{continuity}{momentum}, scalarTransportFoam to determine solute and particle concentrations according to \EQTEXTTWO{solute-dynamics}{particle-dynamics}, and groovyBC (provided by swak4Foam) to implement the boundary conditions in \EQTEXT{boundaries}. We refine the mesh manually in regions where we expect the solute concentration gradient to be large, including the interfaces between inlet streams and near the top and bottom boundaries of the channel. More details about the simulations and mesh design can be found in \APPTEXT{simdetails}.

We consider $\SI{200}{\nano\meter}$ particles in simulations. The P{\'e}clet number associated with particles of diameter $\SI{1}{\micro\meter}$---like those used in experiments with \gls{cps} particles---is large; by considering smaller particles, we artificially enhance the particle diffusion relative to the experiments to improve the stability and convergence of the solver. With larger particles, the particle concentration gradients become prohibitively large and require an extremely fine grid to resolve. This does not affect the qualitative particle focusing dynamics, and the primary consequence is that particle focusing regimes will be more diffuse in simulations than in experiments. The thickness ratio $\NONDIM{\lambda}$ is also affected; the effect is a slightly diminished diffusiophoretic mobility for the $\SI{200}{\nano\meter}$ particles relative to the $\SI{1}{\micro\meter}$ particles used in experiments. Once again, this does not change the qualitative particle dynamics.

\section{Results}
\begin{figure}
    \centering
    \includegraphics{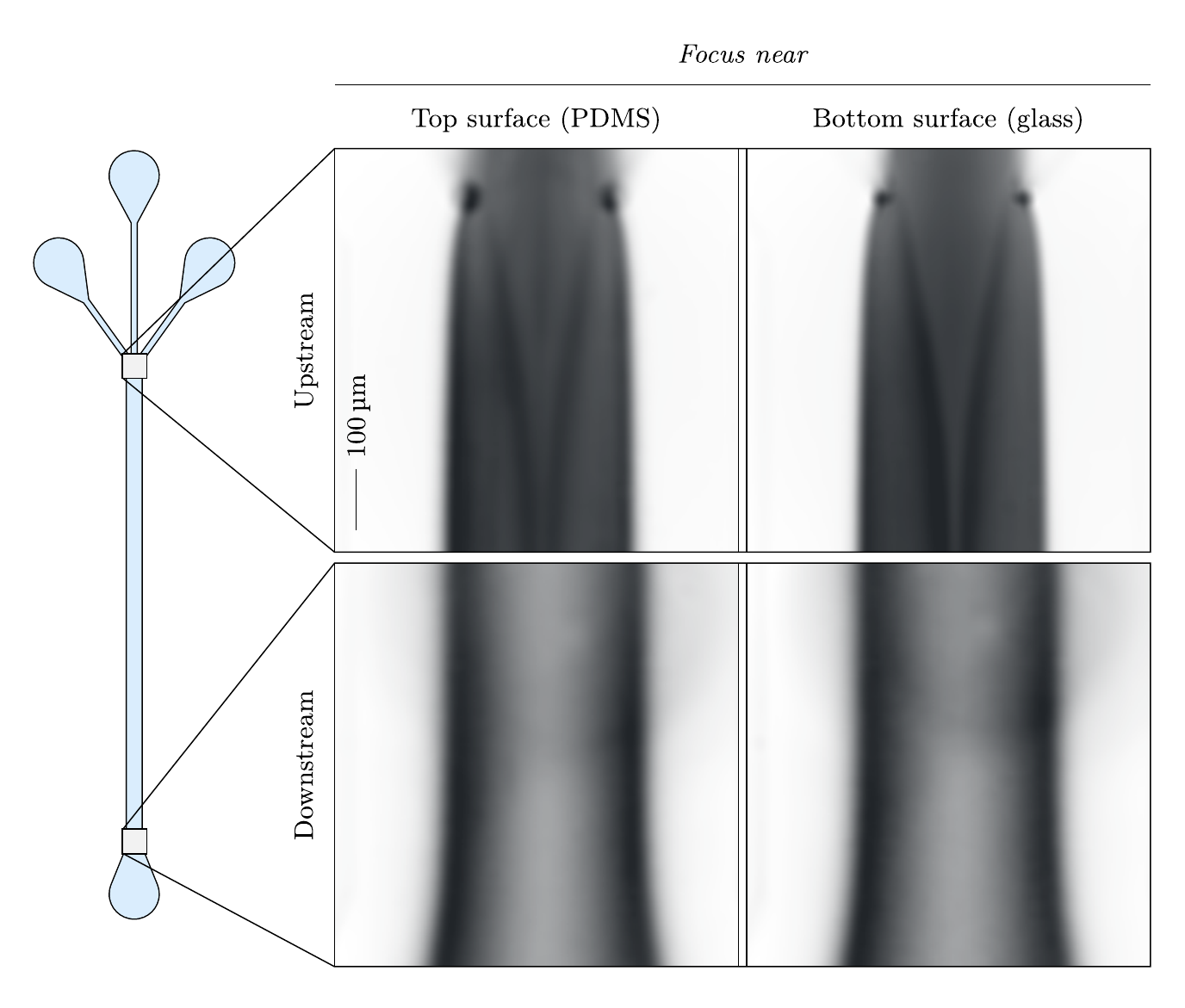}
    \caption{Mean of images (false-color) from experiments for plain \gls{ps} particles with a glass bottom surface. Images are generated with the mean of experimental data and have been smoothed with a Savitsky--Golay filter \cite{Savitzky1964} to reduce the magnitude of deviations caused by particle adhesion and cropped manually. Inward particle migration can be observed in the upstream region. In contrast, the particles downstream do not exhibit this inward migration and instead accumulate only at the outer edge of the particle-rich region.}\label{fig:channel-with-experiments}
\end{figure}

\begin{figure}
    \centering
    \includegraphics{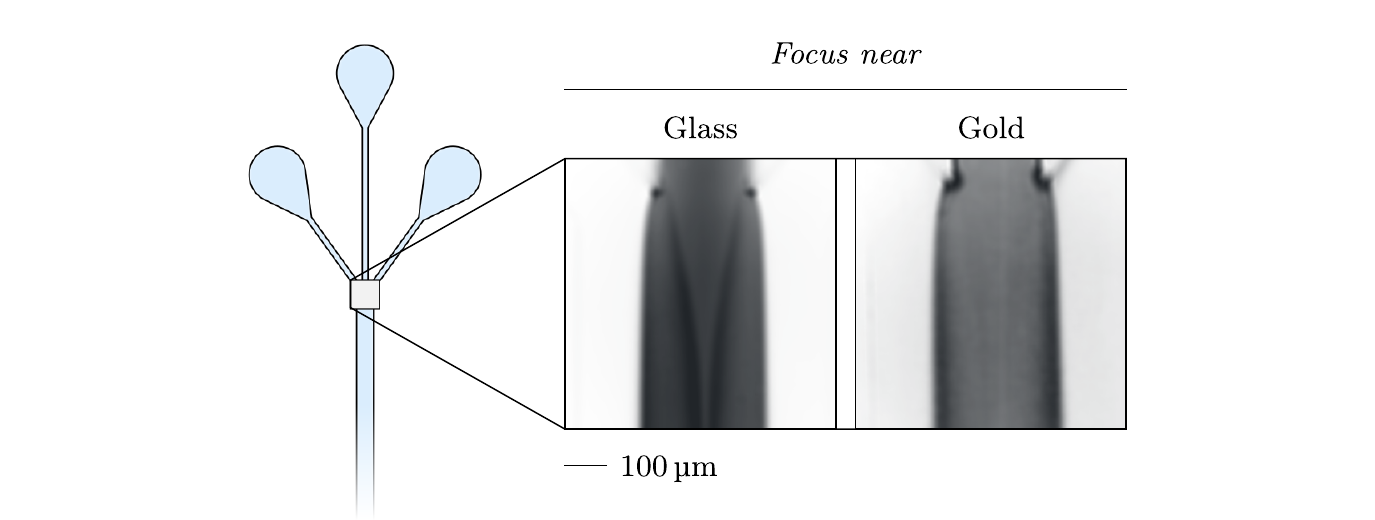}
    \caption{Demonstration of the dependence of the convection rolls structure, visualized with \gls{ps} particles, on the surface material. False-color images are generated with the mean of experimental data and have been smoothed with a Savitsky--Golay filter and cropped manually. The convection rolls, which are visible near glass, vanish when the surface is coated with gold.}\label{fig:glass-gold-comparison}
\end{figure}

\begin{figure}
    \centering
    \includegraphics{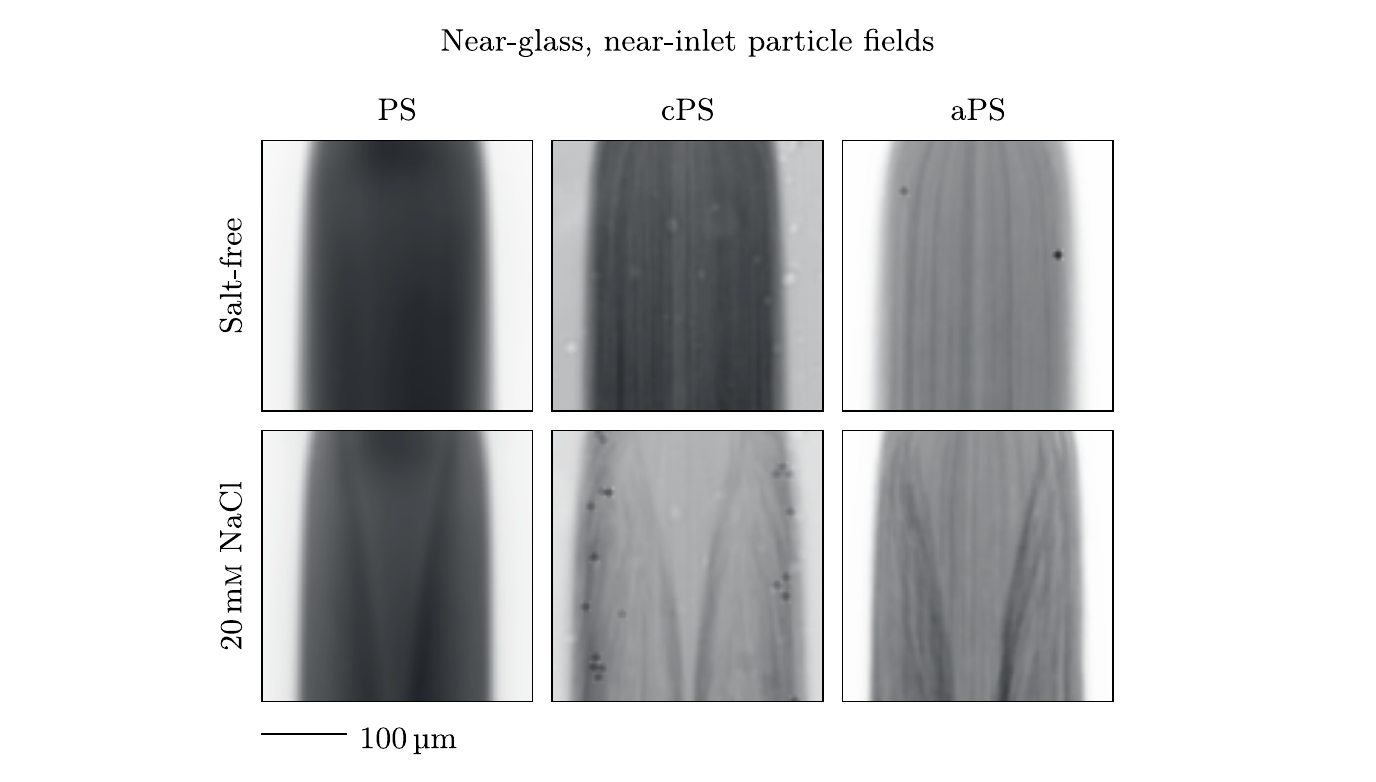}
    \caption{Demonstration of consistency of convection rolls structure in the near-inlet region and its dependence on the existence of a solute concentration gradient. Images (false-color) are generated with the mean of experimental data and have been smoothed with a Savitsky--Golay filter and cropped manually. The qualitative particle focusing behavior is consistent for the different particle species and dependent on the presence of a solute concentration gradient. \gls{ps} particles are $\SI{200}{\nano\meter}$ in size, while \gls{cps} and \gls{aps} particles are $\SI{1}{\micro\meter}$.}\label{fig:particle-species-comparison}
\end{figure}

Near the inlet of the channel, we observe an inward migration of particles toward $y=0$---visible in \FIGTEXT{channel-with-experiments}---near the bottom surface when it is glass; this migration vanishes in the case of the gold surface, as shown in \FIGTEXT{glass-gold-comparison}. We call the focal plane of the camera the ``near-wall'' region; the focus is adjusted manually in each experiment, so this does not correspond to an exact offset from the channel boundaries. The particle motion in this region is qualitatively consistent for \gls{ps}, \gls{cps}, and \gls{aps} particles. We demonstrate this in \FIGTEXT{particle-species-comparison}, which depicts the distinctive inward particle migration for each species in the presence of a solute concentration gradient. In the absence of the concentration gradient, we do not observe the inward migration. The near-surface particle motion, therefore, is likely dependent on the properties of the boundary, and it is consistent with the direction and dependence on solute concentration gradients of diffusioosmotic transport, which are shown in the analytical results presented in \APPTEXT{analytical}.

\begin{figure}
    \centering
    \includegraphics{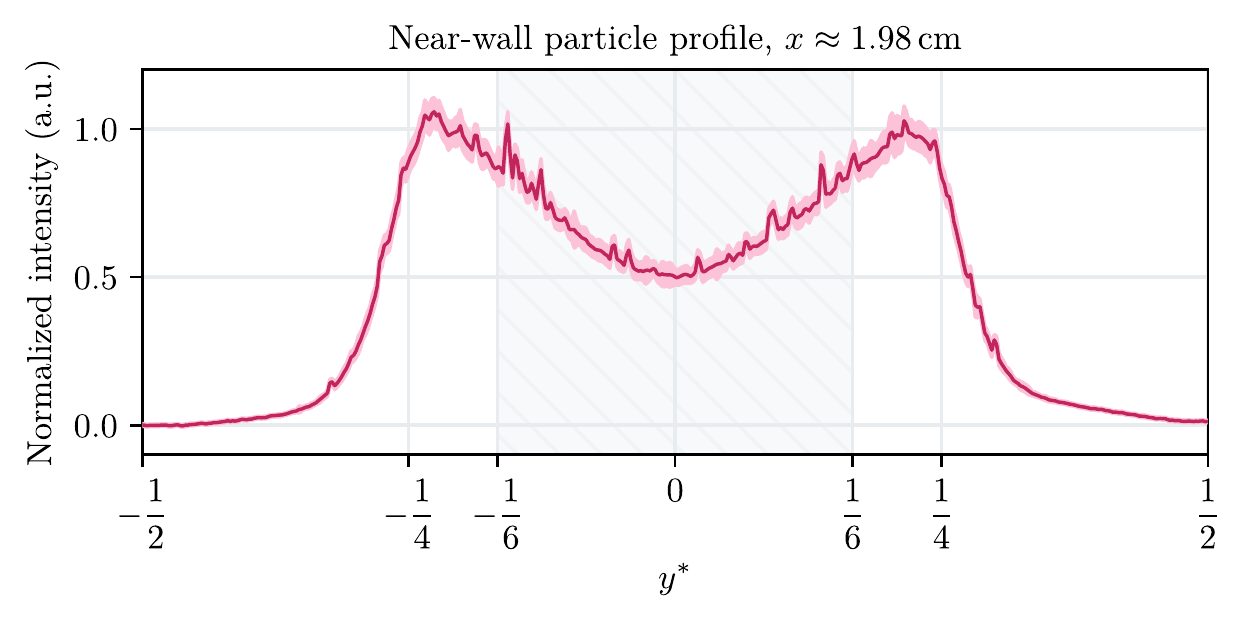}
    \caption{Downstream, near-wall particle concentration profiles at $\DIM{x} \approx \SI{1.98}{\centi\meter}$. The mean is shown as a line and one standard deviation is shaded. The peak particle concentrations occur at $\left| \NONDIM{y} \right| > \frac{1}{6}$ (shaded and hatched), which is characteristic of particle transport driven by diffusiophoresis for this system; we expect particles to migrate outward, toward regions of higher solute concentration. Our results are similar to those given by \citet{Abecassis2008} (\SEE{\FIGTEXTDUMMY{3}}), with particle migration outward from the center of the channel. There are minor differences attributable to near-wall confinement effects and differences in species, geometry, and flow conditions.}\label{fig:profile-glass-downstream}
\end{figure}

\begin{figure}
    \centering
    \includegraphics{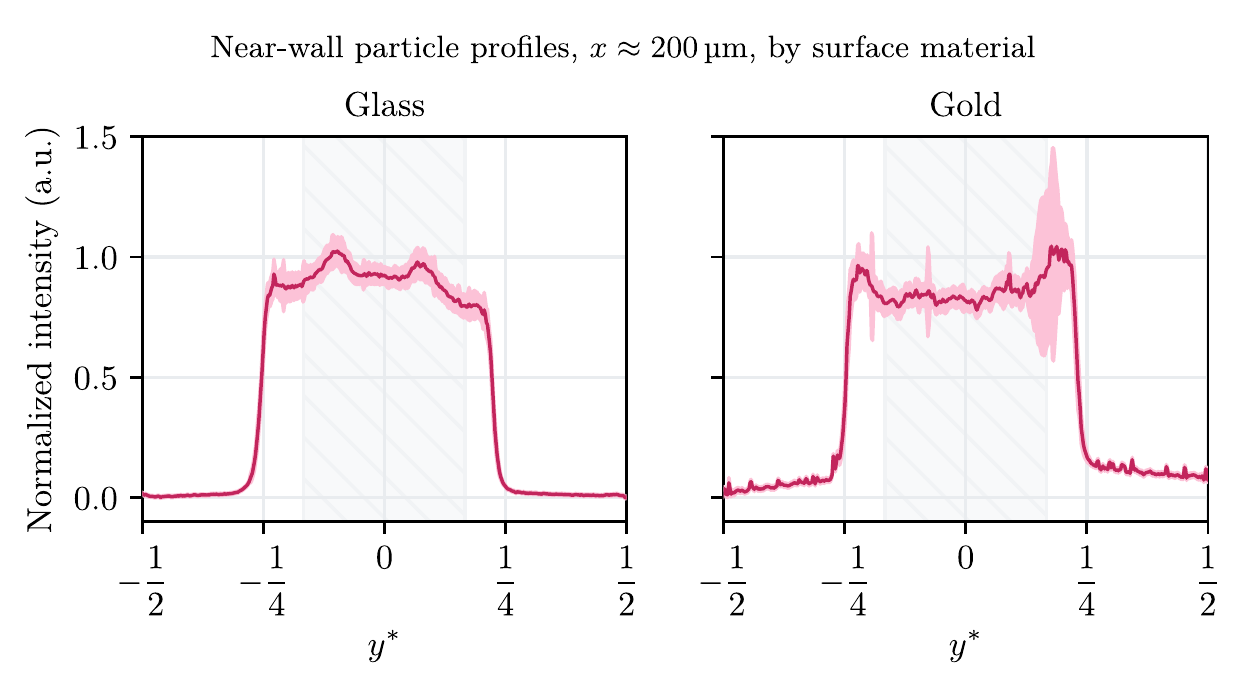}
    \caption{Upstream, near-wall particle concentration profiles at $\DIM{x} \approx \SI{200}{\micro\meter}$. The mean is shown as a line and one standard deviation is shaded. Our results near the gold surface are similar to those given by \citet{Abecassis2008} (\SEE{\FIGTEXTDUMMY{3}}), but the profile near the glass surface---with peak particle concentrations at $\left| \NONDIM{y} \right| < \frac{1}{6}$ without a change in solute or particle species or locations---is not seen by \citet{Abecassis2008}. This migration of the location of the peak particle concentration is consistent with recent results from \citet{Chakra2023}. Near the gold surface, the particle concentration peaks are at $\left| \NONDIM{y} \right| > \frac{1}{6}$, as in \FIGTEXT{profile-glass-downstream}. Near the glass surface, however, the peak concentration occurs at $\left| \NONDIM{y} \right| < \frac{1}{6}$, which suggests particle migration is directed inward.}\label{fig:particle-profile-comparison}
\end{figure}

Our experimental results for near-wall concentration profiles are qualitatively consistent with published results \cite{Abecassis2008} downstream, as shown in \FIGTEXT{profile-glass-downstream}, and upstream in the case of the gold surface, as shown in \FIGTEXT{particle-profile-comparison}. There are slight differences that result from the proximity to the wall, but the distinctive change in the location of the peak particle concentration, which we associate with diffusioosmotic flow, is not seen near the gold surface or downstream near the glass surface. The upstream concentration profiles near the glass surface, however, are distinct from those previously reported by \citet{Abecassis2008}; the peak particle concentration near the glass surface is found at a position $\left| \NONDIM{y} \right| < \frac{1}{6}$, which indicates that particles migrate inward in the near-wall, near-inlet region. This is consistent with the results of recent work by \citet{Chakra2023}. Notably, there is considerably more noise in the results near the gold surface, which is likely a result of particle adhesion. It is likely that adhesion is not as significant near glass because both the \gls{ps} particles and glass surface are negatively charged.

\begin{figure}
    \centering
    \includegraphics{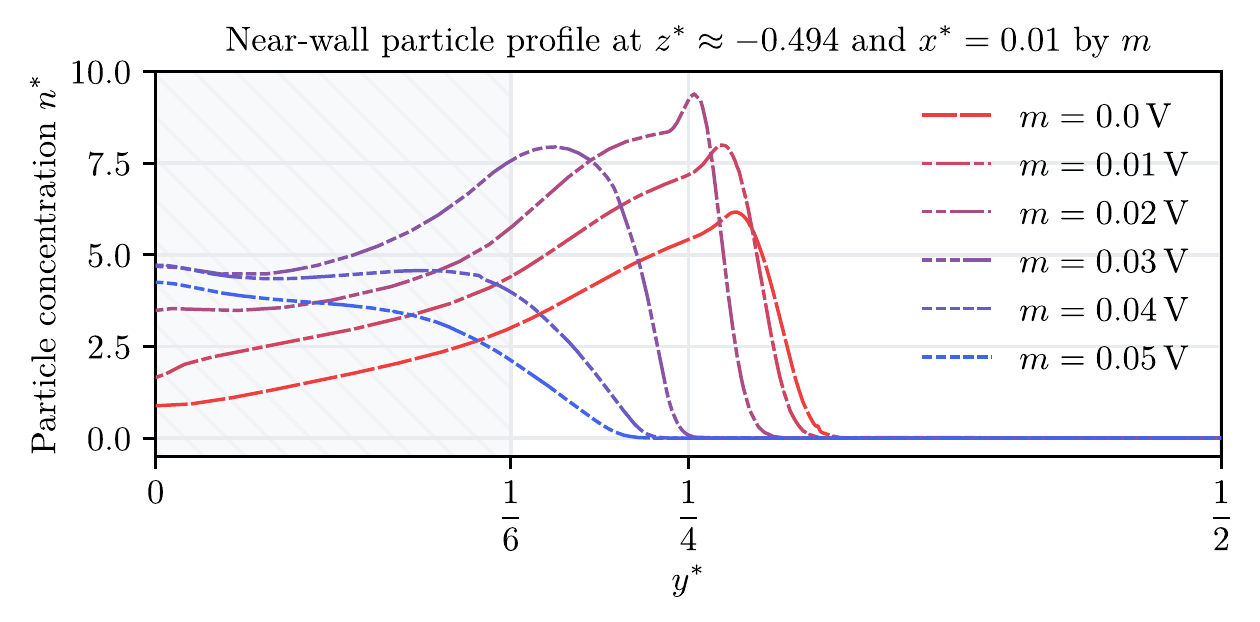}
    \caption{Simulated particle concentration profiles at $\NONDIM{z} \approx -0.494$ and $\NONDIM{x}=0.01$ for increasing magnitudes of the diffusioosmotic mobility. The outward migration of particles is most significant at $\DIMCONST{m}=\SI{0}{\volt}$, in the absence of diffusioosmosis. At larger $\DIMCONST{m}$, the peak particle concentration moves toward $\NONDIM{y}=0$. The particle concentration in the center of the channel is also affected: Solute is advected toward the center of the channel, where it is drawn upward, along the charged surface. The resulting plume of solute yields a concentration gradient toward the center of the channel, affecting the particle dynamics in turn. We show the effect of this plume on the particle concentration field in two dimensions in \FIGTEXT{wall-focusing}.}\label{fig:sim-profiles}
\end{figure}

We provide numerical estimates for the near-wall particle profiles, analogous to \FIGTEXTTWO{profile-glass-downstream}{particle-profile-comparison}, in \FIGTEXT{sim-profiles}. Direct comparison between the experiments and simulations is not possible because of the uncertainty in the diffusioosmotic mobility and differences in particle size; the consistency of the experimental results in the near-inlet region, however, allows us to generalize and comment about trends and features rather than trying to match an experimental case exactly. The effect of increasing the zeta potential and diffusioosmotic mobility by increasing $\DIMCONST{m}$ in $\DIM{\zeta_{\rm b}} = \DIMCONST{m} \log_{10} \NONDIM{c}$ (see \APPTEXT{zetamobilitymodels}) is to move the location of the peak particle concentration inward toward $\NONDIM{y}=0$. Another consequence is that there is focusing near the wall throughout the cross-section: $\NONDIM{n} \left( \NONDIM{y} = 0 \right)$ increases with the value of $\DIMCONST{m}$. This occurs because the diffusioosmotic flow draws solute inward along the surface of the channel, which establishes a concentration gradient toward the boundary. As the magnitude of the diffusioosmotic mobility is increased further, a third particle concentration peak forms in the center of the channel because the solute entrained by the flow along the bottom surface is advected away from the wall as it approaches $\NONDIM{y}=0$.

\begin{figure}
    \centering
    \includegraphics{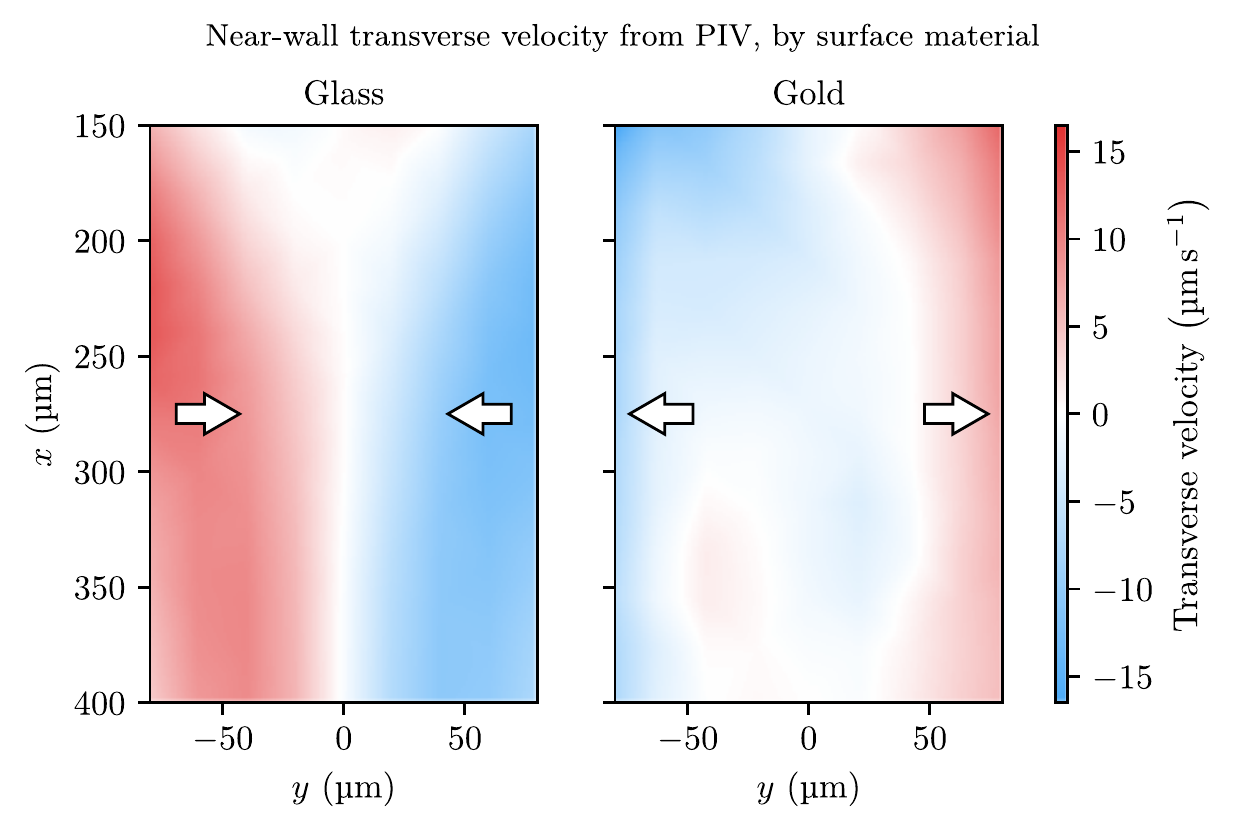}
    \caption{Transverse particle velocity in the $\DIM{y}$-direction from experimental images, obtained with \gls{piv}. The structure of the convection rolls is clearly visible near the glass surface; particles are drawn inward toward $\NONDIM{y}=0$. This structure vanishes near the gold surface, and the direction of net particle migration reverses. This reversal of the direction of particle migration is observed in simulations in \FIGTEXT{velocity-profiles-sim}.}\label{fig:velocity-profiles-exp}
\end{figure}

\begin{figure}
    \centering
    \includegraphics{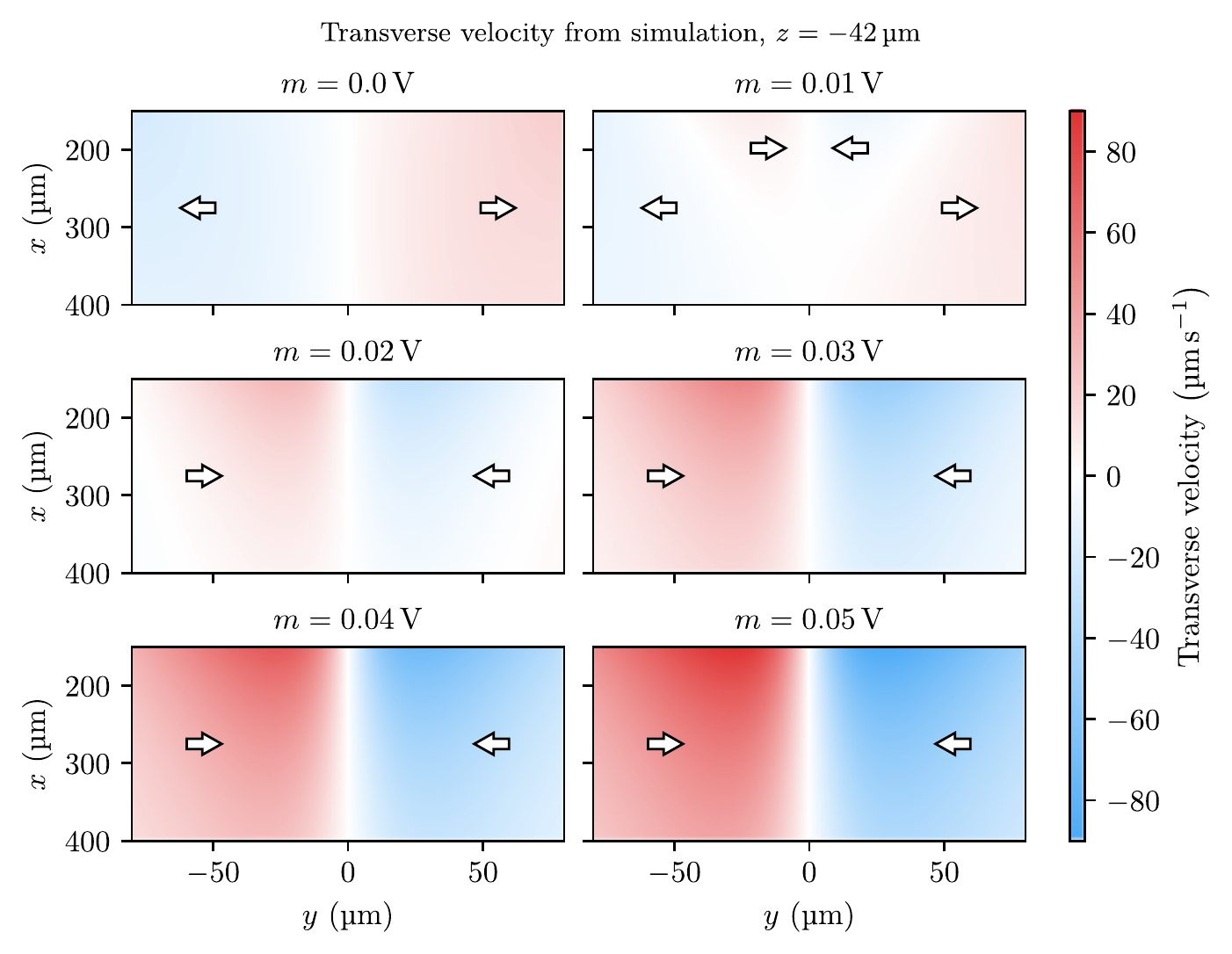}
    \caption{Transverse particle velocity in the $\DIM{y}$-direction from simulations with varying $\DIMCONST{m}$. When $\DIMCONST{m}=\SI{0}{\volt}$, particles migrate outward from $\NONDIM{z}=0$ because of the background velocity profile and diffusiophoresis. As $\DIMCONST{m}$ increases, particle focusing at the center of the channel becomes more significant. This reversal of the direction of particle migration is observed experimentally in \FIGTEXT{velocity-profiles-exp}. Notably, the dynamics near the channel inlet are distinct and the diffusioosmotic velocity is greater in the near-inlet region in simulations than in experiments.}\label{fig:velocity-profiles-sim}
\end{figure}

The convection rolls are also apparent in the velocity component in the $\DIM{y}$-direction in our experimental results and simulations. In \FIGTEXT{velocity-profiles-exp}, we show the near-wall velocity profile from \gls{piv} near the glass or gold surface. Near the glass surface, particles are drawn toward $\NONDIM{y}=0$; this structure vanishes near the gold surface and the direction of net particle migration is instead outward from the center of the channel. Our simulations, shown in \FIGTEXT{velocity-profiles-sim} for $\DIM{z}=\SI{-42}{\micro\meter}$, capture this behavior qualitatively, though they overpredict the particle velocity near the inlet, which is apparent when comparing \FIGTEXTTWO{velocity-profiles-exp}{velocity-profiles-sim}. Where $\DIMCONST{m}=\SI{0}{\volt}$, particles migrate outward from $\NONDIM{y}=0$ because of the combined influence of the background flow and diffusiophoresis, but the direction of motion reverses when the diffusioosmotic velocity is sufficiently large to balance this outward migration.

\begin{figure}
    \centering
    \includegraphics{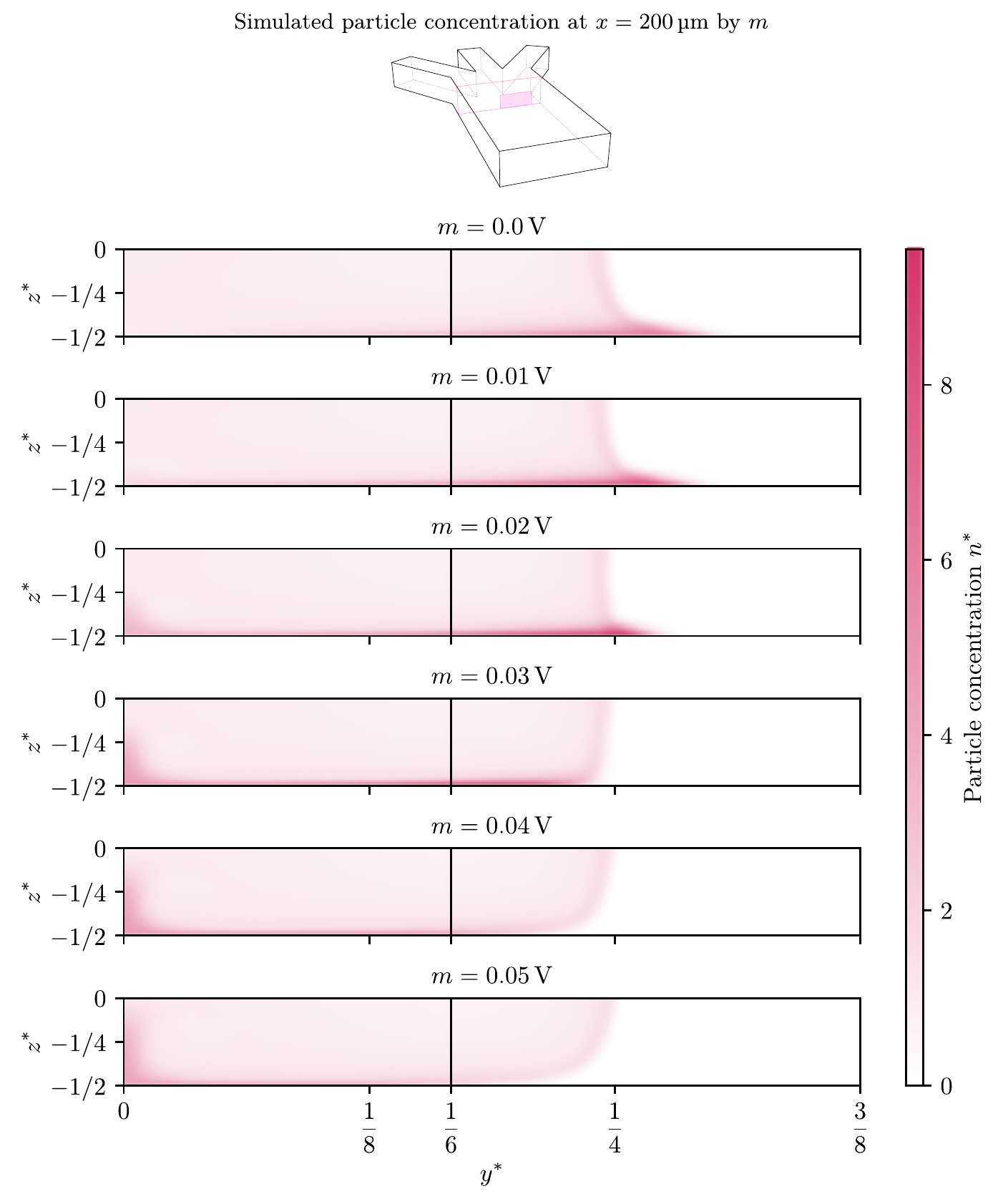}
    \caption{Slices of particle concentration at $\DIM{x} = \SI{200}{\micro\meter}$ (shaded on diagram). We show a line at $\NONDIM{y}=\frac{1}{6}$ to denote the region where particles are introduced to the channel. They have moved outward through diffusion and diffusiophoresis, and because the transverse component of $\DIM{\VEC{\VEC{u}}}$ is nonzero. As $\DIMCONST{m}$ increases, the zeta potential and diffusioosmotic mobility of the wall increase in magnitude. This causes particles near the wall to be drawn toward the center of the channel at a greater velocity. Additionally, the slip flow at $\NONDIM{z}=-\frac{1}{2}$ draws solute from the solute-rich regions at $\NONDIM{y} > \frac{1}{6}$ toward $\NONDIM{y}=0$, where it is advected upward by the convection rolls. The resulting solute concentration gradient yields a plume of particles, which migrate through diffusiophoresis, at $\NONDIM{y}=0$.}\label{fig:wall-focusing}
\end{figure}

\begin{figure}
    \centering
    \includegraphics{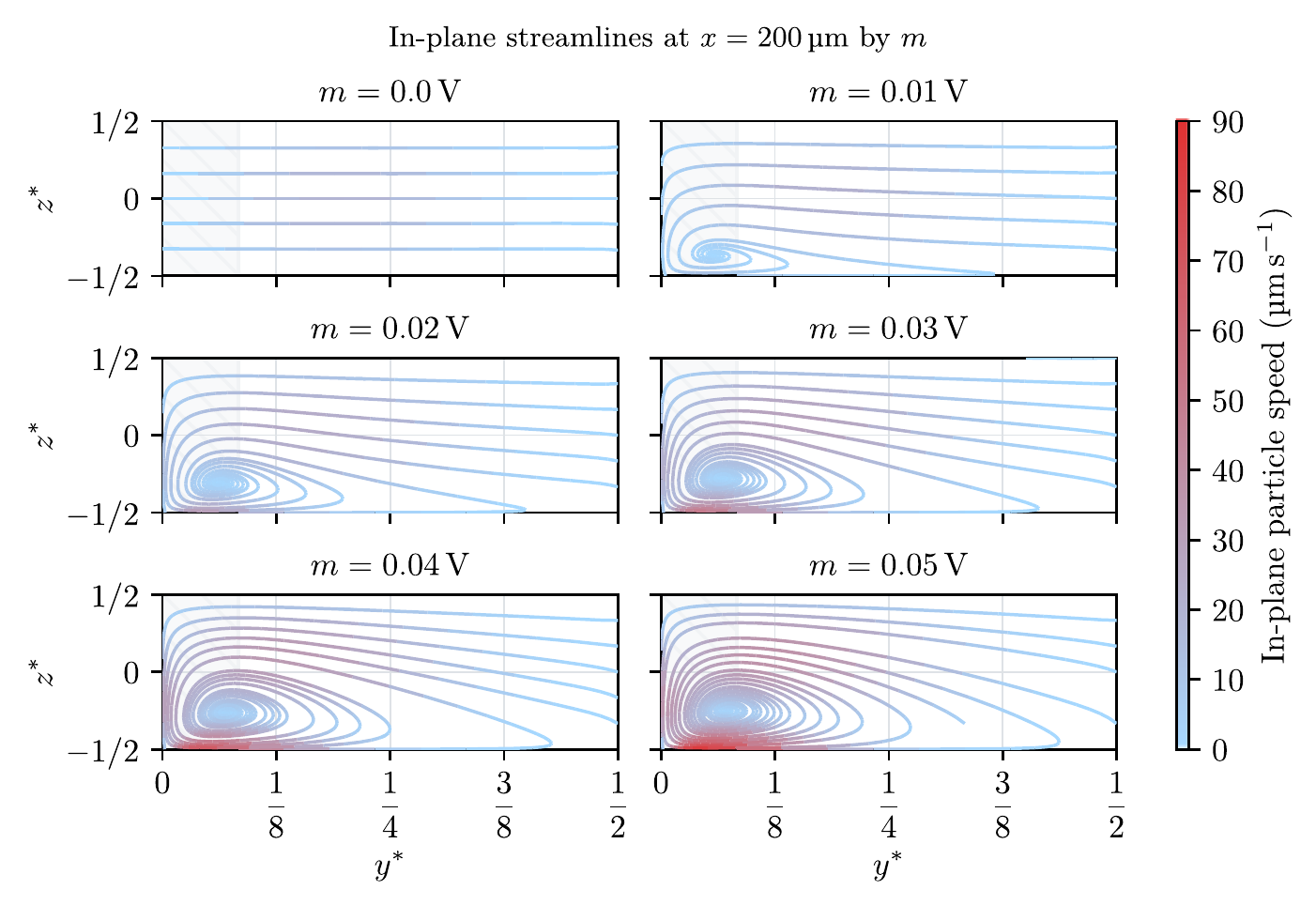}
    \caption{Streamlines at a slice $\DIM{x}=\SI{200}{\micro\meter}$ by $\DIMCONST{m}$. The particle speed as shown includes both the background flow and diffusiophoresis. The sense of the convection rolls is clockwise in all images. We have neglected the out-of-plane velocity to better show the shape of the convection rolls. As $\DIMCONST{m}$ increases, the particle velocity increases and the convection rolls grow in size. The scale of the convection rolls can be significant relative to the channel depth when the surface is highly charged.}\label{fig:roll-velocity}
\end{figure}

We observe significant particle focusing at channel walls, even in the absence of diffusioosmosis. This focusing occurs because the solute concentration profile near walls is more diffuse than in the center of the channel, where the streamwise velocity is larger; that is, solute near boundaries has longer to diffuse than solute in the center of the channel before reaching the same position in the $\DIM{x}$-direction. As a result, there is a nonzero component of the solute concentration gradient---and, consequently, of the diffusiophoretic velocity---that is directed toward the upper and lower walls. We demonstrate that the focusing is significant in \FIGTEXT{wall-focusing}, which shows the particle concentration is largest at the wall, even where diffusioosmosis is neglected (i.e., the case where $\DIMCONST{m} = \SI{0}{\volt}$). This can also be observed in analytical results given in \APPTEXT{analytical}, where particles migrate in the $z$-direction even in cases where the diffusioosmotic mobility is zero. The convection rolls, shown in \FIGTEXT{roll-velocity}, draw particles along the wall, where they are advected upward in the center of the channel. This behavior is also captured by the first-order model given in \APPTEXT{analytical}. The strength of this effect increases with the magnitude of $\DIMCONST{m}$, which sets the diffusioosmotic mobility.

\section{Discussion}
Particle focusing toward the upper or lower walls can be significant even when the surfaces are uncharged. We have previously commented on particle accrual near walls and corners through diffusiophoresis in three-dimensional geometries \cite{Migacz2022}, but we considered the deformation of a plug of solute rather than merging streams with distinct concentrations. We observe the same phenomenon where streams of distinct concentration merge. We demonstrate focusing near walls in \FIGTEXT{wall-focusing}; in a system where $\DIM{{\mathcal{M}}} < 0$, this focusing would instead be at the center of the channel. Notably, in our system, the focusing of particles at channel walls is enhanced by the contrast between inlet flow rates. The flow is faster in the center of the channel because the flow rates to \FIGTEXT{channel}(A) and (B) are both $\SI{1}{\micro\liter\per\minute}$, but the former is split across two inlets. Consequently, there is a transverse velocity outward from $\NONDIM{y}=0$ exclusively due to the background fluid flow. This enhances the component of the concentration gradient directed toward the boundary. \citet{Chakra2023} make a similar observation about the diffusiophoretic migration of particles toward boundaries. It may be possible to reverse the direction of focusing or enhance the effect by simply changing the inlet flow rates in our system.

The three-dimensional nature of the particle dynamics in channel flows is highly relevant to mixing in microfluidic processes. The implications of diffusiophoresis on mixing at the microscale have been examined in previous works; \citet{Deseigne2014} comment, for example, on the impact of solutes on the mixing of colloids in a channel with a staggered herringbone pattern. Our work demonstrates that diffusioosmotic flow can have a significant impact on particle dynamics in smooth channels, absent obstacles or patterns, provided solute concentration gradients are present at boundaries of nonzero surface charge. The choice of materials for microfluidic devices can impact the dynamics of both solutes and particles in the presence of solute concentration gradients. Materials with relatively high surface charge could be selected to maximize this mixing effect; materials with low surface charge could be used to minimize it. The approximate solution for the fluid, solute, and particle dynamics, given in \APPTEXT{analytical}, can be used to estimate the effect of diffusioosmosis and diffusiophoresis on solute and particle species without significant computational resources or experiments; it can be readily adapted to work with other boundary conditions. The analytical solution provides a good approximation to the dynamics of the convection rolls, which is apparent from the similarity between \FIGTEXTTWO{velocity-profiles-exp}{velocity-profiles-sim} and \FIGTEXT{analytical-top-velocity} in the appendix.

We neglected diffusioosmotic flow at the top of the channel in simulations because we were interested in the dynamics near the bottom boundary, which is the only surface for which we considered different materials in experiments. The \gls{pdms} has a nonzero surface charge, however, and convection rolls develop near the top of the channel in all experiments; this is readily apparent in \FIGTEXT{channel-with-experiments}, in which the particle concentration peaks near the center of the channel can be seen near both the top and bottom surfaces in the near-inlet region. Notably, diffusioosmotic flow at the side walls of the channel is negligible because there is not a significant variation in solute concentration near those walls. Considering the effects of multiple boundaries with nonzero charge could be an interesting area of further research. Another avenue of further study is to consider additional particle species with distinct diffusiophoretic mobility. The \gls{ps} species we use may have similar diffusiophoretic mobilities, which is apparent in the comparison of mobility estimates for various \gls{ps} species in \ce{NaCl} gradients shown in the appendix in \FIGTEXT{mobility-comparison}. Indeed, \FIGTEXTSTART{exp-mobility} shows that particle migration at low concentrations is in the same direction, and of similar magnitude, for our \gls{ps}, \gls{cps}, and \gls{aps} particles. Particles with $\DIM{{\mathcal{M}}} < 0$ would migrate in the same direction through both diffusiophoresis and the effect of diffusioosmosis in this system, which would likely enhance the transverse velocity and increase the strength of the convection rolls. Another potentially interesting area of further research is the interaction of the diffusioosmosis-driven particle focusing with phenomena such as the focusing of particles in the presence of a surfactant gradient and complexing polymer, recently explored by \citet{Yang2023}.

Empirical models for particle properties may contribute to uncertainty in our results; the models for zeta potential as a function of $\ln \NONDIM{c}$, for instance, have linear asymptotes and are unbounded. Extrapolation to concentrations not considered with zeta potentiometry could yield erroneous results for particle migration. This is particularly relevant at low concentrations, where solution conductivity is minimal and measurement of electrophoretic mobility through conventional means is difficult. Additionally, the model for the zeta potential of the particles is only appropriate when the Debye length is small relative to the particle size \cite{KirbyBook}, which is a further source of uncertainty at low solute concentrations. We have also neglected variations in other quantities. One example of such a quantity is the pH, which may be affected by the intrusion of \ce{CO2} or other gaseous species through the gas-permeable \gls{pdms} walls \cite{Shim2021,Shim2020,Shin2017b,Shimokusu2019}. This is unlikely to affect the dynamics near the glass or gold surfaces but may warrant further study for other materials commonly used in microfluidic devices. We have also neglected variation in density and other fluid properties along solute concentration gradients. \citet{Gu2018} and \citet{Williams2020} show that such variations can result in buoyancy-driven flows with a magnitude that is dependent on channel geometry and solute species. They consider more significant concentration gradients, however, and variations in the properties of \ce{NaCl} solutions at or below approximately $\SI{20}{\milli\Molar}$ are negligible. This is apparent from the near symmetry across the $\DIM{x}\DIM{y}$-plane in \FIGTEXT{channel-with-experiments}; the direction of particle migration is the same on the top and bottom surface, which we expect to have similar surface charges, and this symmetry would be broken if convection were the dominant transport mechanism (\SEE{\FIGTEXTDUMMY{2}}~of~\cite{Williams2020}).

Our experimental observations support the use of models of variable diffusiophoretic and diffusioosmotic mobilities. To demonstrate the importance of variable-mobility models, we plot an effective near-wall mobility in \FIGTEXT{exp-mobility}, noting that both diffusioosmosis and diffusiophoresis have a logarithmic dependence on the solute concentration gradient and are additive very close to the wall. The plot shows the quotient of $\left( \DIM{\VEC{\VEC{u}}_{\text{PIV}}} - \DIM{\VEC{\VEC{u}}} \right) \cdot \UNITVEC{e}_{\DIM{y}}$ and $\left( \DIMCONST{\VEC{\nabla}} \ln \NONDIM{c} \right) \cdot \UNITVEC{e}_{\DIM{y}}$, where $\UNITVEC{e}_{\DIM{y}}$ is the unit vector in the $\DIM{y}$-direction. Here, we find the velocity $\DIM{\VEC{\VEC{u}}_{\text{PIV}}}$ from the experimental results with \gls{piv}, while we determine $\DIM{\VEC{u}}$ and $\NONDIM{c}$ from a simulation without diffusioosmosis. This provides a rough estimate, provided the effect of diffusioosmosis on the solute concentration field is small and the focal plane of the microscope is close to the channel surface, for the effective mobility, which includes the effects of both diffusiophoresis and diffusioosmosis. As we expect, we observe outward particle migration near the gold surface. Notably, the particles move inward in the case of the glass surface, but there are visible inflection points where the effective mobility begins to decrease in magnitude. In the case of the $\SI{1}{\micro\meter}$ \gls{cps} particles, the mobility changes sign within the concentration range we consider, indicating that the motion of particles near the surface is now directed outward from $\NONDIM{y}=0$. The dynamics of the particles in the inlet and outlet regions, therefore, can be distinct; a change in the magnitude of the solute concentration can have a significant impact on near-wall velocity, even when the direction of the gradient is constant, as a result of concentration-dependent diffusiophoretic and diffusioosmotic mobilities. This supports the conclusions of \citet{Akdeniz2023}, who have recently found that models of zeta potential as a function of solute concentration are important to accurately describe particle motion in a pore over long times.

\begin{figure}
    \centering
    \includegraphics{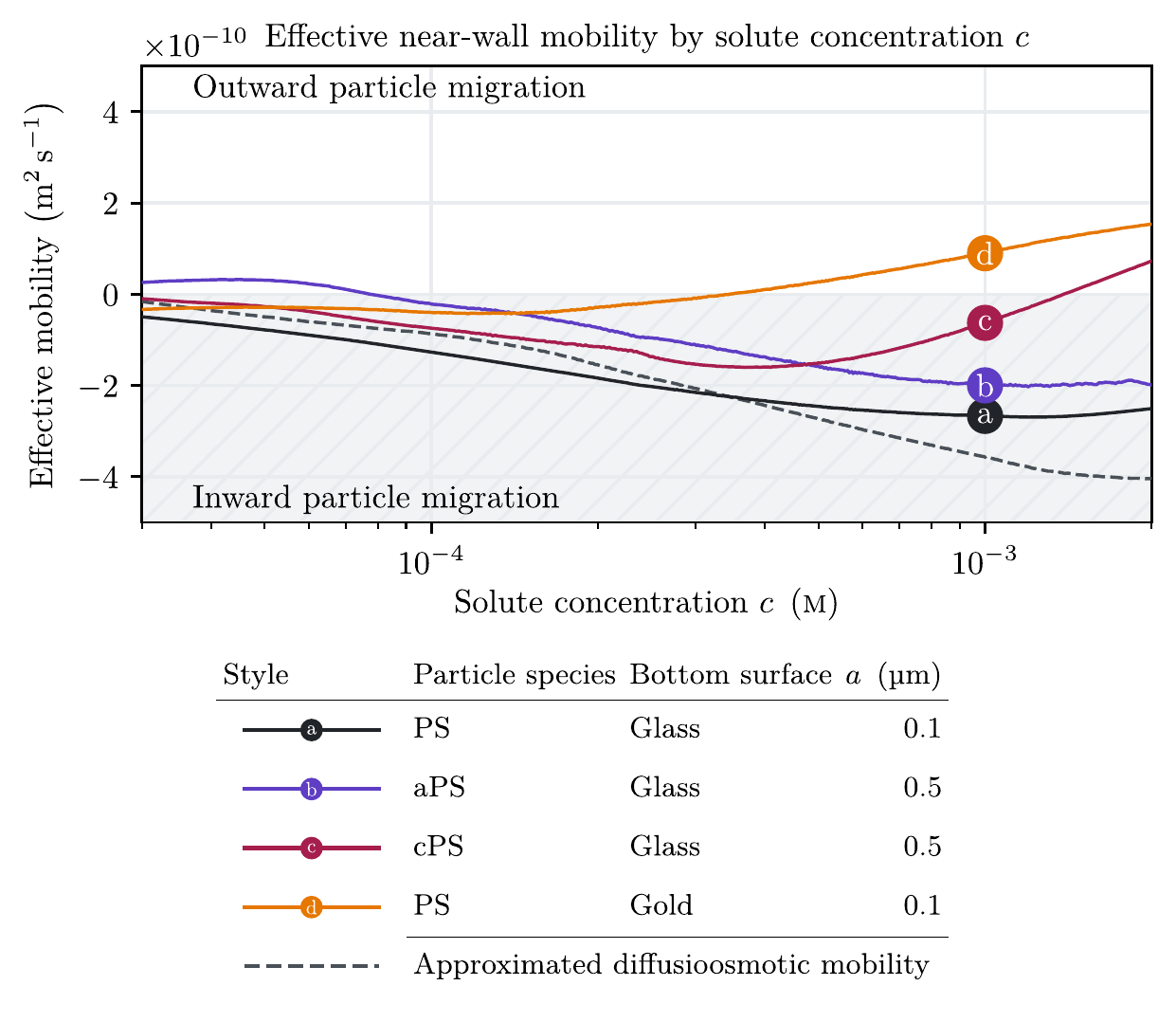}
    \caption{Approximate net near-wall mobility. To obtain an estimate for the net mobility, we compare experimental velocities and numerical values of $\DIMCONST{\VEC{\nabla}} \ln \NONDIM{c}$---interpolated to the same $100 ts 100$ grid with $\DIM{x} \in \left[\SI{100}{\micro\meter}, \SI{300}{\micro\meter} \right]$ and $\DIM{y} \in \left[\SI{10}{\micro\meter}, \SI{180}{\micro\meter} \right]$---in the absence of diffusioosmosis. We show a moving average over $100$ values. The approximated diffusioosmotic mobility is the difference between the net mobilities of \gls{ps} particles with glass and gold boundaries, denoted (a) and (d).}\label{fig:exp-mobility}
\end{figure}

Our experimental results for the effective mobility contradict the model we use for the wall zeta potential and diffusioosmotic mobility using published data for silica. With the model we use for the wall zeta potential, the mobility is unbounded and continues to increase as the solute concentration is lowered. We observe that the effective mobility diminishes in magnitude at low concentration, and because the diffusiophoretic mobility decreases as $\NONDIM{c}$ is lowered (see \FIGTEXTTWO{zeta-concentration}{mobility-concentration} or the near-wall mobility in the case of the gold surface in \FIGTEXT{exp-mobility}), the experimental results are consistent only with a similar decay in diffusioosmotic mobility at low concentrations. This is also apparent when comparing \FIGTEXTTWO{velocity-profiles-exp}{velocity-profiles-sim}; the behavior in the near-inlet region, where the solute concentration is lowest, is not described accurately by the model, which otherwise provides estimates for velocity that are both qualitatively consistent with experiments and appropriate in magnitude. This difference could be attributable, in part, to differences in surface chemistry arising from either properties of the material itself or surface treatments such as plasma cleaning. Indeed, the diffusioosmotic mobility is a significant source of uncertainty in our calculations, and estimation is further complicated by electrokinetic lift \cite{Wu1996}, which may also account for the lack of a third particle peak at $\NONDIM{y}=0$ in \FIGTEXT{channel-with-experiments} when it is seen in simulated particle profiles like those shown in \FIGTEXT{sim-profiles}. The velocity in the $\DIM{z}$-direction would not be as significant at the center of the channel if the diffusioosmotic velocity were to decay at low solute concentrations. The surface zeta potential may also be time-dependent, which is something we have not considered; for some materials, such as \gls{pdms} \cite{Makamba2003}, surface fouling over time may be an important consideration. The diffusiophoretic mobility, however, is likely reasonable: \citet{Keh2000} comment that the model agrees well with previously published results ``up to [a zeta potential $\DIM{\zeta}$ of] $\SI{50}{\milli\volt}$,'' which is consistent with the order of magnitude of $\DIM{\zeta}$ we consider (see \FIGTEXT{zeta-concentration} in the appendix).

\section{Conclusion}
We have described the three-dimensional dynamics of solute and particles in merging streams of distinct solute concentration experimentally, numerically, and analytically. Near walls of nonzero surface charge, diffusioosmosis results in the migration of particles along the boundary; this does not occur near an uncharged surface. This phenomenon could have implications for microfluidic devices for which mixing processes are relevant, as near-wall flows can be exploited to enhance or suppress mixing. Additionally, we contend that the change from inward to outward particle migration we observe near the boundary can be described by solute concentration-dependent models of diffusiophoretic and diffusioosmotic mobility, which is direct evidence to support recent trends toward the adoption of variable-mobility models in studies of diffusiophoresis and diffusioosmosis.

\section{Acknowledgments}
We gratefully acknowledge Viet Sang Doan and Sangwoo Shin at the University at Buffalo for performing experiments to quantify the particle zeta potential. Part of this research was conducted using computational resources and services at the Center for Computation and Visualization, Brown University.

\appendix

\section{Approximate solution\label{sec:app:analytical}}
\begin{figure}
    \centering
    \includegraphics{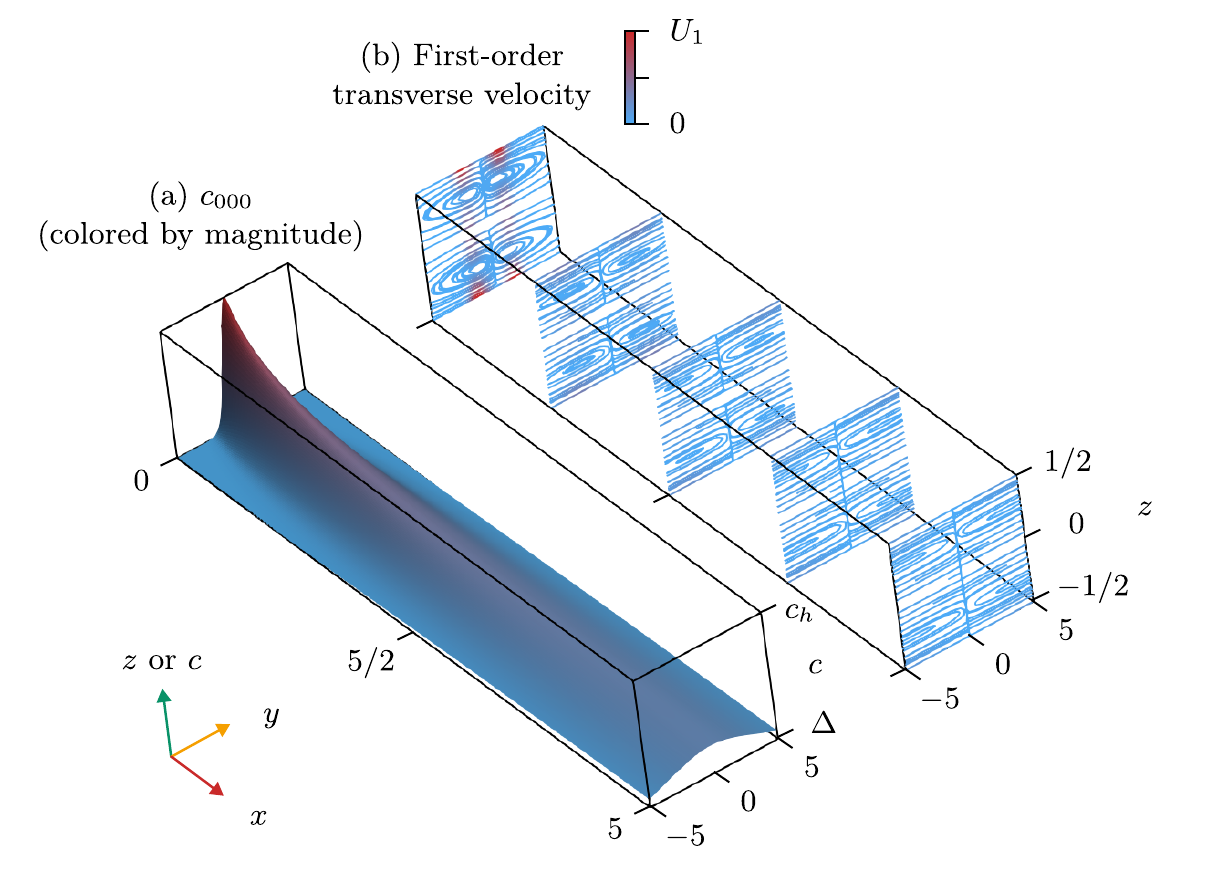}
    \caption{Example of (a) zeroth-order solute concentration and (b) streamlines of first-order transverse velocity for a case where $\Delta < 1$ and $\alpha_p=0$. The concentration is bounded by $\Delta$ and $c_h = \left(1 - \Delta\right) / \sqrt{\pi} + \Delta$. The maximum particle speed $U_1$ is affected by $\alpha$ and $\Delta$. In cases where $\Delta > 1$, the concentration at $y=0$ is lower than the concentration as $y \rightarrow \infty$ and the direction of motion is reversed relative to the system depicted here.}\label{fig:3d-appendix}
\end{figure}

Note that all variables in this section are dimensionless; the notation differs from the main text. We study the convection rolls phenomenon in a simplified system, where the initial concentration profile is Gaussian. This allows us to consider the solute concentration in a similarity regime, unlike cases where the initial solute profile is a step, as in the main text. The depth of the channel in the $z$-direction is $\ell_3$; the $x$- and $y$-dimensions are infinite. The problem we consider here is similar to the numerical work of \citet{Chakra2023}, though the inlet solute concentration and the semi-infinite channel geometry are distinct. An example of the system is shown in \FIGTEXT{3d-appendix} with the zeroth-order solute concentration and first-order transverse velocity.

The dimensionless continuity equation is
\begin{equation}
\frac{\partial u}{\partial x} + \frac{\partial v}{\partial y} + \frac{\partial w}{\partial z} = 0,
\end{equation}
where we have used a characteristic scale for velocity $U$ in the $x$-direction, $V=\frac{{\mathcal{D}}_c}{\ell_2}$ in the $y$-direction, and $W=\frac{{\mathcal{D}}_c \ell_3}{\ell_2^2}$ in the $z$-direction. The flow in the channel is governed by the Stokes equation, which gives
\begin{align}
0 &= -\frac{p_c}{\ell_1} \frac{\partial p}{\partial x} + \mu \left[ \frac{{\mathcal{D}}_c}{\ell_1 \ell_2^2} \frac{\partial^2 u}{\partial x^2} + \frac{U}{\ell_2^2} \frac{\partial^2 u}{\partial y^2} + \frac{U}{\ell_3^2} \frac{\partial^2 u}{\partial z^2}\right],\\
0 &= -\frac{p_c}{\ell_2} \frac{\partial p}{\partial y} + \mu \left[ \frac{{\mathcal{D}}_c}{\ell_1 \ell_2^2} \frac{\partial^2 v}{\partial x^2} + \frac{{\mathcal{D}}_c}{\ell_2^3} \frac{\partial^2 v}{\partial y^2} + \frac{{\mathcal{D}}_c}{\ell_2 \ell_3^2} \frac{\partial^2 v}{\partial z^2}\right],\quad\text{and}\\
0 &= -\frac{p_c}{\ell_3} \frac{\partial p}{\partial z} + \mu \left[ \frac{{\mathcal{D}}_c \ell_3}{\ell_1^2 \ell_2^2} \frac{\partial^2 w}{\partial x^2} + \frac{{\mathcal{D}}_c \ell_3}{\ell_2^4} \frac{\partial^2 w}{\partial y^2} + \frac{{\mathcal{D}}_c}{\ell_2^2 \ell_3} \frac{\partial^2 w}{\partial z^2}\right]
\end{align}
in the $x$-, $y$- and $z$-directions, respectively. Here, $p_c$ is a characteristic pressure. The length scale in the $y$-direction, $\ell_2$, is fixed by the width of the inlet solute profile. The value $\ell_1$, which provides a length scale in the $x$-direction, is related to $\ell_2$ through the diffusion equation. Intuitively, we expect that the dominant terms are the pressure gradient in the $x$-direction and the viscous $\frac{\partial^2 u}{\partial z^2}$ term, which must balance. Therefore, we choose a pressure scale $p_c = \frac{\mu U \ell_1}{\ell_3^2}$, which gives
\begin{align}
0 &= -\frac{\partial p}{\partial x} + \frac{\epsilon^2}{\text{Pe}} \frac{\partial^2 u}{\partial x^2} + \epsilon^2 \frac{\partial^2 u}{\partial y^2} + \frac{\partial^2 u}{\partial z^2},\\
0 &= -\text{Pe} \frac{\partial p}{\partial y} + \frac{\epsilon^2}{\text{Pe}} \frac{\partial^2 v}{\partial x^2} + \epsilon^2 \frac{\partial^2 v}{\partial y^2} + \frac{\partial^2 v}{\partial z^2},\quad\text{and}\\
0 &= -\text{Pe} \frac{\partial p}{\partial z} + \frac{\epsilon^4}{\text{Pe}} \frac{\partial^2 w}{\partial x^2} + \epsilon^4 \frac{\partial^2 w}{\partial y^2} + \epsilon^2 \frac{\partial^2 w}{\partial z^2},
\end{align}
where $\text{Pe}=\frac{U\ell_1}{{\mathcal{D}}_c}$ is the P{\'e}clet number and $\epsilon = \frac{\ell_3}{\ell_2}$ is the channel aspect ratio. At the walls at $z=\pm \frac{1}{2}$, we impose slip boundary conditions to account for diffusioosmosis; these are
\begin{equation}
u\left( x, y, \pm \frac{1}{2} \right) = -\frac{1}{\text{Pe}} \frac{\mathcal{M}_{{{\text{do}}}}}{{\mathcal{D}}_c} \frac{\partial \ln c}{\partial x},\quad
v\left( x, y, \pm \frac{1}{2} \right) = -\frac{\mathcal{M}_{{{\text{do}}}}}{{\mathcal{D}}_c} \frac{\partial \ln c}{\partial y},\quad\text{and}\quad w\left(x, y, \pm \frac{1}{2} \right) = 0.
\end{equation}
The boundary conditions can be modified to account for different channel materials. At the inlet of the channel, the velocity is
\begin{equation}
u(0, y, z) = \frac{3}{2} \left( 1 - 4 z^2 \right),
\quad
v(0, y, z) = 0,
\quad\text{and}\quad
w(0, y, z) = 0.
\end{equation}
The dimensionless advection--diffusion equation for the solute is
\begin{equation}
\frac{U}{\ell_1} u \frac{\partial c}{\partial x} + \frac{V}{\ell_2} v \frac{\partial c}{\partial y} + \frac{W}{\ell_3} w \frac{\partial c}{\partial z}
=
\frac{{\mathcal{D}}_c}{\ell_1^2} \frac{\partial^2 c}{\partial x^2} + \frac{{\mathcal{D}}_c}{\ell_2^2} \frac{\partial^2 c}{\partial y^2} + \frac{{\mathcal{D}}_c}{\ell_3^2} \frac{\partial^2 c}{\partial z^2},
\end{equation}
which we simplify to write
\begin{equation}
\epsilon^2 \left( u \frac{\partial c}{\partial x} + v \frac{\partial c}{\partial y} + w \frac{\partial c}{\partial z} \right) = \frac{\epsilon^2}{\text{Pe}} \frac{\partial^2 c}{\partial x^2} + \epsilon^2 \frac{\partial^2 c}{\partial y^2} + \frac{\partial^2 c}{\partial z^2}.
\end{equation}

We consider small parameters $\alpha = \frac{\mathcal{M}_{{{\text{do}}}}}{{\mathcal{D}}_c}$, $\beta = \frac{1}{\rm Pe} = \frac{{\mathcal{D}}_c}{U \ell_1}$, $\epsilon^2 = \frac{\ell_3^2}{\ell_2^2}$, and $\alpha_p = \frac{\mathcal{M}}{{\mathcal{D}}_c}$. To simplify notation, we use $A_{ijk}$ to denote the term of the expansion of $A$ at $\mathcal{O} (\alpha^i \beta^j \epsilon^{2k})$. This gives the series
\begin{equation}
\begin{aligned}
A &= A_{000}+\epsilon^2 A_{001}+\epsilon^4 A_{002}+\beta A_{010}\\&\hphantom{=}+\beta\epsilon^2 A_{011}+\beta^2 A_{020}+\alpha A_{100}+\alpha\epsilon^2 A_{101}\\&\hphantom{=}+\alpha\beta A_{110}+\alpha^2 A_{200}+ \ldots
\end{aligned}
\end{equation}
for each variable $A$. As a result of the one-way coupling of the particle dynamics to the solute dynamics, the velocity, pressure, and solute concentration are independent of $\alpha_p$. We use the similarity solution to the one-dimensional diffusion equation in an infinite domain, where we use position $x$ as a substitute for time, for the leading-order solute profile. This is
\begin{equation}\label{eq:c000}
c_{000} = \frac{1 - \Delta}{\sqrt{4 \pi x + \pi}} \exp \left( - \frac{y^2}{4x + 1} \right) + \Delta,
\end{equation}
where $\Delta$ sets the solute contrast. This is similar to solutions for the solute concentration given by \citet{TengDiffusioosmosis} and \citet{Gupta2020a}. When $\Delta > 1$, the particle concentration is largest at $\left| y \right| \rightarrow \infty$ and smallest at $y=0$; when $\Delta < 1$, the concentration is largest at $y=0$.

\subsection{Solute dynamics\label{sec:analyticalsolute}}
The solute concentration profile and the velocity profile are coupled as a result of diffusioosmosis. The leading-order velocity and pressure are, from Poiseuille flow,
\begin{align}
u_{000} &= \frac{3}{2} \left( 1 - 4 z^2 \right),\\
v_{000} &= 0,\\
w_{000} &= 0,\quad\text{and}\\
p_{000} &= -12 x.
\end{align}
Deviation from this background flow is a result of diffusioosmosis at boundaries. We first note that the $z$-momentum equation at $\mathcal{O}(\alpha)$ yields $\frac{\partial p_{100}}{\partial z} = 0$, so $p_{100}$ is independent of $z$. To leading order, the governing equations are now
\begin{align}
\frac{\partial u_{100}}{\partial x} + \frac{\partial v_{100}}{\partial y} + \frac{\partial w_{100}}{\partial z} &= 0\quad\text{(continuity),}\\
-\frac{\partial p_{100}}{\partial x} + \frac{\partial^2 u_{100}}{\partial z^2} &= 0\quad\text{($x$-momentum),}\label{eq:ymomentum0}\\
\frac{\partial p_{100}}{\partial y} &= 0\quad\text{($y$-momentum),}\label{eq:pressure100}\\
\frac{\partial p_{101}}{\partial z} &= 0\quad\text{($z$-momentum), and}\label{eq:pressure101}\\
-\frac{\partial^2 c_{000}}{\partial y^2}
+ \frac{3 \left( 1 - 4 z^2 \right)}{2} \frac{\partial c_{000}}{\partial x}
- \frac{\partial^2 c_{001}}{\partial z^2}
&=
0\quad{\text{(advection--diffusion)}}.\label{eq:advectiondiffusion0}
\end{align}
\EQTEXTTWOSTART{pressure100}{pressure101} suggest that $p_{100}$ is a function only of $x$ and $p_{101}$ is a function of $x$ and $y$. At the next order, the $y$- and $z$-momentum equations are, respectively,
\begin{align}
\frac{\partial p_{110}}{\partial y} - \frac{\partial^2 v_{100}}{\partial z^2} &= 0\quad\text{and}\\
\frac{\partial p_{111}}{\partial z} - \frac{\partial^2 w_{100}}{\partial z^2} &= 0.
\end{align}
\EQTEXTSTART{ymomentum0} can be solved to yield
\begin{equation}
u_{100} = k_1 + k_2 z + \frac{z^2}{2} \frac{\dd p_{100}}{\dd x},
\end{equation}
where $k_1$ and $k_2$ are fixed by the leading-order boundary conditions,
\begin{align}
k_1 + \frac{k_2}{2} + \frac{1}{8} \frac{\dd p_{100}}{\dd x} &= 0\quad\text{and}\label{eq:bc1}\\
k_1 - \frac{k_2}{2} + \frac{1}{8} \frac{\dd p_{100}}{\dd x} &= 0.\label{eq:bc2}
\end{align}
\EQTEXTTWOSTART{bc1}{bc2} yield
\begin{align}
k_1 &= -\frac{1}{8} \left( 4 k_2 - \frac{\dd p_{100}}{\dd x} \right)\quad\text{and}\\
k_2 &= 0.
\end{align}
The correction to the velocity in the $x$-direction is then
\begin{equation}
u_{100} = \frac{4z^2 - 1}{8} \frac{\dd p_{100}}{\dd x};
\end{equation}
note, however, that the integral of higher-order terms over a cross-section must be zero to conserve mass. This requires constant $p_{100}$. Consequently, the velocity correction is $u_{100}=0$, and the first-order correction to the velocity field from diffusioosmotic flow includes only transverse components. The $y$-momentum equation is now, to leading order,
\begin{equation}\label{eq:ymomentum1}
\frac{\partial p_{110}}{\partial y} - \frac{\partial^2 v_{100}}{\partial z^2} = 0.
\end{equation}
Solving \EQTEXT{ymomentum1} yields
\begin{equation}
v_{100} = j_1 + j_2 z + \frac{z^2}{2} \frac{\partial p_{110}}{\partial y};
\end{equation}
solution in the same manner as the last order gives the functions
\begin{align}
j_1 &= -\frac{j_2}{2} - \frac{1}{c_{000}} \frac{\partial c_{000}}{\partial y} - \frac{1}{8} \frac{\partial p_{110}}{\partial y}\quad\text{and}\\
j_2 &= 0,
\end{align}
so
\begin{equation}
v_{100}=-\frac{1}{c_{000}} \frac{\partial c_{000}}{\partial y} + \frac{4z^2 - 1}{8} \frac{\partial p_{110}}{\partial y}.
\end{equation}
The continuity equation is now
\begin{equation}
\frac{8}{c_{000}} \left( \frac{\partial c_{000}}{\partial y} \right)^2 - 8 \frac{\partial^2 c_{000}}{\partial y^2}
+ c_{000} \left[
\left( 4z^2 - 1 \right) \frac{\partial^2 p_{110}}{\partial y^2} + 8 \frac{\partial w_{100}}{\partial z}
\right]
= 0
\end{equation}
and can be solved to find
\begin{equation}
w_{100} = r + \frac{1}{8 c_{000}^2} \left[
-8 z \left( \frac{\partial c_{000}}{\partial y} \right)^2
- \frac{4}{3} z^3 c_{000}^2 \frac{\partial^2 p_{110}}{\partial y^2}
+ z c_{000} \left( 8 \frac{\partial^2 c_{000}}{\partial y^2} + c_{000} \frac{\partial^2 p_{110}}{\partial y^2} \right)
\right].
\end{equation}
Applying the boundary condition at $z = \frac{1}{2}$, we find
\begin{equation}
r = \frac{1}{24 c_{000}^2} \left[
12 \left( \frac{\partial c_{000}}{\partial y} \right)^2
- 12 c_{000} \frac{\partial^2 c_{000}}{\partial y^2}
- c_{000}^2 \frac{\partial^2 p_{110}}{\partial y^2}
\right];
\end{equation}
similarly, applying the boundary condition at $z = -\frac{1}{2}$ yields
\begin{equation}
p_{110} = h_1 + h_2 y - 12 \ln c_{000}.
\end{equation}
To conserve mass, $h_2$ must be zero. The velocity components are now
\begin{align}
v_{100} &= \frac{1 - 12z^2}{2 c_{000}} \frac{\partial c_{000}}{\partial y}\quad\text{and}\\
w_{100} &= \frac{4z^3 - z}{2 c_{000}^2} \left[ - \left( \frac{\partial c_{000}}{\partial y} \right)^2 + c_{000} \frac{\partial^2 c_{000}}{\partial y^2} \right].
\end{align}
\EQTEXTSTART{advectiondiffusion0} can be solved to find
\begin{equation}\label{eq:c001}
c_{001} = m_1 + m_2 z - \frac{z^2}{2} \frac{\partial^2 c_{000}}{\partial y^2} + \frac{3 z^2}{4} \frac{\partial c_{000}}{\partial x} - \frac{z^4}{2} \frac{\partial c_{000}}{\partial x}
\end{equation}
with $m_2 = 0$ from the no-flux boundary condition. The function $m_1$ does not affect the leading-order particle dynamics.

\begin{figure}
    \centering
    \includegraphics{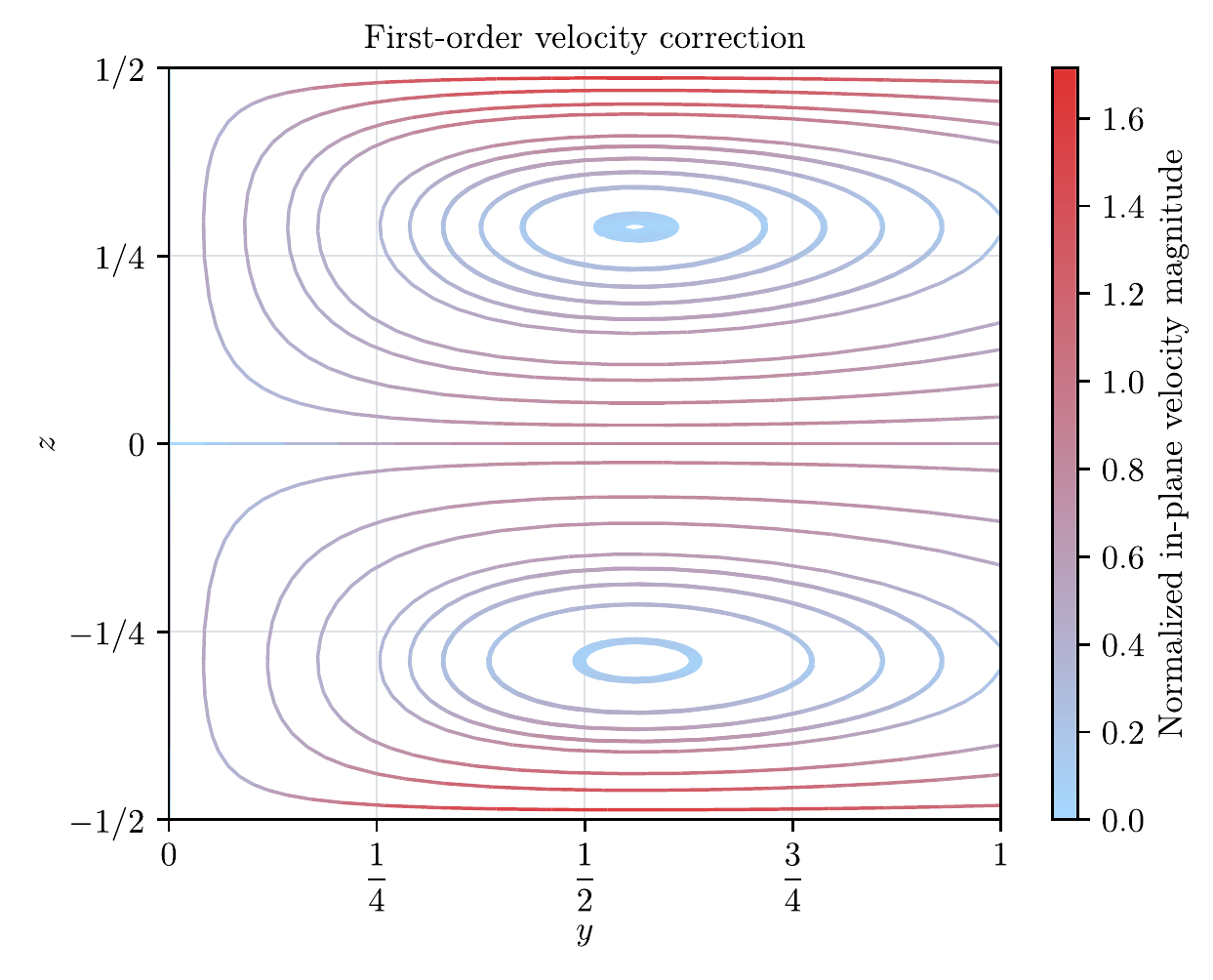}
    \caption{Example of first-order velocity correction at $x=0$ with $\Delta = 10$. The convection rolls draw fluid toward $y=0$ at the boundaries when $\alpha > 0$; the velocity is largest at the boundaries, where diffusioosmosis occurs. The velocity profile is symmetric about $y=0$ and is similar to profiles observed in numerical simulations. The presence of a second roll at $z=\frac{1}{2}$ is a result of the fact that we consider a case with uniform diffusioosmotic mobilities at both the upper and lower walls.}\label{fig:firstordervelocity}
\end{figure}

\begin{figure}
    \centering
    \includegraphics{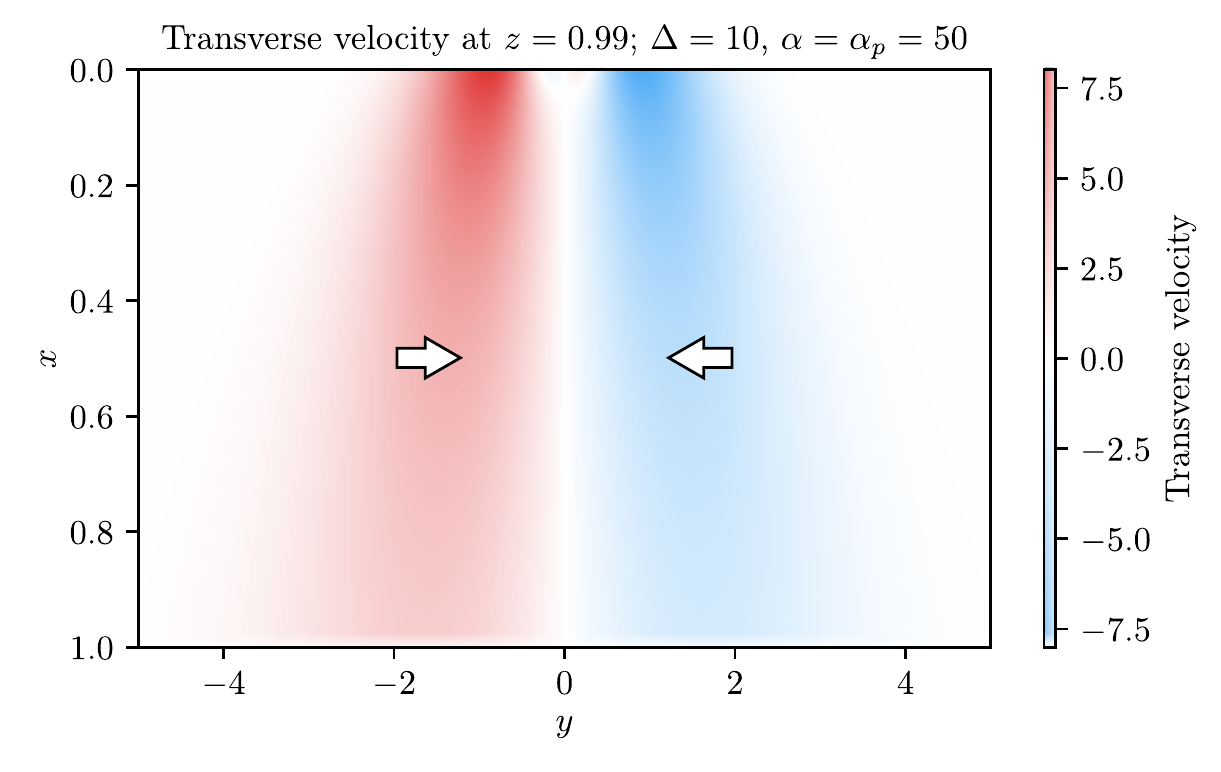}
    \caption{Velocity in the $y$-direction at $z=0.99$ with $\Delta=10$ and $\alpha=\alpha_p=50$. The velocity profile is similar to that seen in \FIGTEXTTWO{velocity-profiles-exp}{velocity-profiles-sim}, which indicates the solution is capturing the relevant dynamics. Solute and particles near the boundary are drawn inward toward $y=0$.}\label{fig:analytical-top-velocity}
\end{figure}

The corrections $u_{100}$, $v_{100}$, and $w_{100}$ give the first-order velocity profile that results from diffusioosmosis at the boundaries of the channel and allow for the visualization of the structure of the convection rolls. We show an example in \FIGTEXT{firstordervelocity}, which shows the characteristic structure of the convection rolls with the largest velocity at the boundary. In the example, particles are drawn inward toward $y=0$ along the wall, where they are advected away from the wall and are drawn outward by the recirculating flow. We show another example of the structure in \FIGTEXT{analytical-top-velocity}, which is reminiscent of \FIGTEXTTWO{velocity-profiles-exp}{velocity-profiles-sim} from experiments and simulations, respectively. It shows the transverse velocity that draws solute and particles toward the center of the channel near the boundary.

\subsection{Particle dynamics}
\begin{figure}
    \centering
    \includegraphics{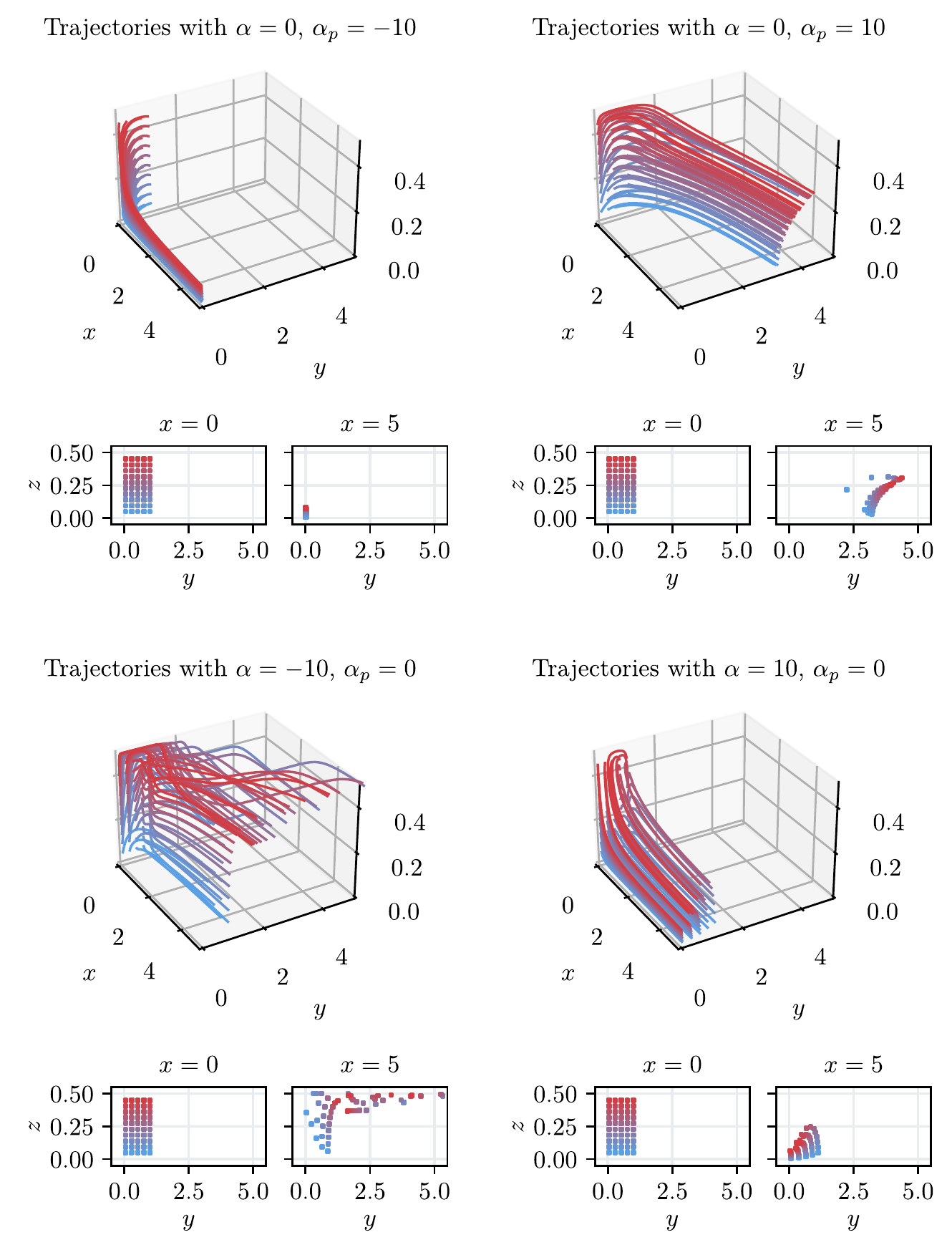}
    \caption{Example particle trajectories with large values of $\alpha$ and $\alpha_p$. Here, $\Delta=10$ and the solute concentration is lowest in the center of the channel. The theory is valid for small parameters, but we use large values to demonstrate the effects of diffusiophoresis and diffusioosmosis on particle trajectories. When $\alpha_p < 0$, the particles migrate toward the center of the channel, where the solute concentration is lowest; when $\alpha_p > 0$, the particles migrate outward toward higher solute concentration. The particles migrate in the $z$-direction because of the effects of walls on the solute concentration profile. When $\alpha < 0$, particles are drawn outward along the wall because of diffusioosmosis; they are drawn toward the center of the channel along the wall when $\alpha > 0$. Particle paths are colored by initial $z$-position.}\label{fig:trajectoriesmultiple}
\end{figure}

To visualize the effects of diffusiophoresis and diffusioosmosis on the particle dynamics, we calculate particle trajectories for individual particles with a position $\VEC{x}_p = \left( x_p, y_p, z_p \right)$. The nondimensional particle position is governed by
\begin{align}
\frac{\dd x_p}{\dd t} &= u + \alpha_p\beta \frac{\dd \ln c}{\dd x},\\
\frac{\dd y_p}{\dd t} &= v + \alpha_p \frac{\dd \ln c}{\dd y},\quad\text{and}\\
\frac{\dd z_p}{\dd t} &= w + \frac{\alpha_p}{\epsilon^2} \frac{\dd \ln c}{\dd z};
\end{align}
to first order, this is
\begin{align}
\frac{\dd x_p}{\dd t} &= u_{000},\\
\frac{\dd y_p}{\dd t} &= \alpha v_{100} + \frac{\alpha_p}{c_{000}} \frac{\partial c_{000}}{\partial y},\quad\text{and}\\
\frac{\dd z_p}{\dd t} &= \alpha w_{100} + \frac{\alpha_p}{c_{000}} \frac{\partial c_{001}}{\partial z}.
\end{align}
We show example trajectories for large $\alpha$ and $\alpha_p$ in \FIGTEXT{trajectoriesmultiple}. While the theory is strictly valid for small parameters, here we use large $\alpha$ and $\alpha_p$ so the trajectories demonstrate the effects of diffusiophoresis and diffusioosmosis on particles. Large values of $\alpha$ and $\alpha_p$ are necessary to visualize the structures because the concentration gradient is not steep, as it is in the near-inlet region of the channel, with the initial concentration profile defined in \EQTEXT{c000}. In this example, $\Delta = 10$, which corresponds to an initial solute profile that is largest at large $y$ and smallest at $y=0$. When $\alpha_p < 0$, particles migrate down the solute concentration gradient and focus in the center of the channel. When $\alpha_p > 0$, they migrate up the solute concentration gradient, away from $y=0$. Notably, the particles migrate in the $z$-direction even in the absence of diffusioosmosis because the presence of walls affects the solute profile. When $\alpha < 0$, diffusioosmosis draws particles outward from $y=0$ along the wall. Particles near $z=0$ migrate toward $y=0$ because of the recirculation of the fluid. The opposite occurs when $\alpha > 0$: particles at the wall are drawn inward toward $y=0$, while particles near $z=0$ are drawn outward by the recirculating flow.

\begin{figure}
    \centering
    \includegraphics{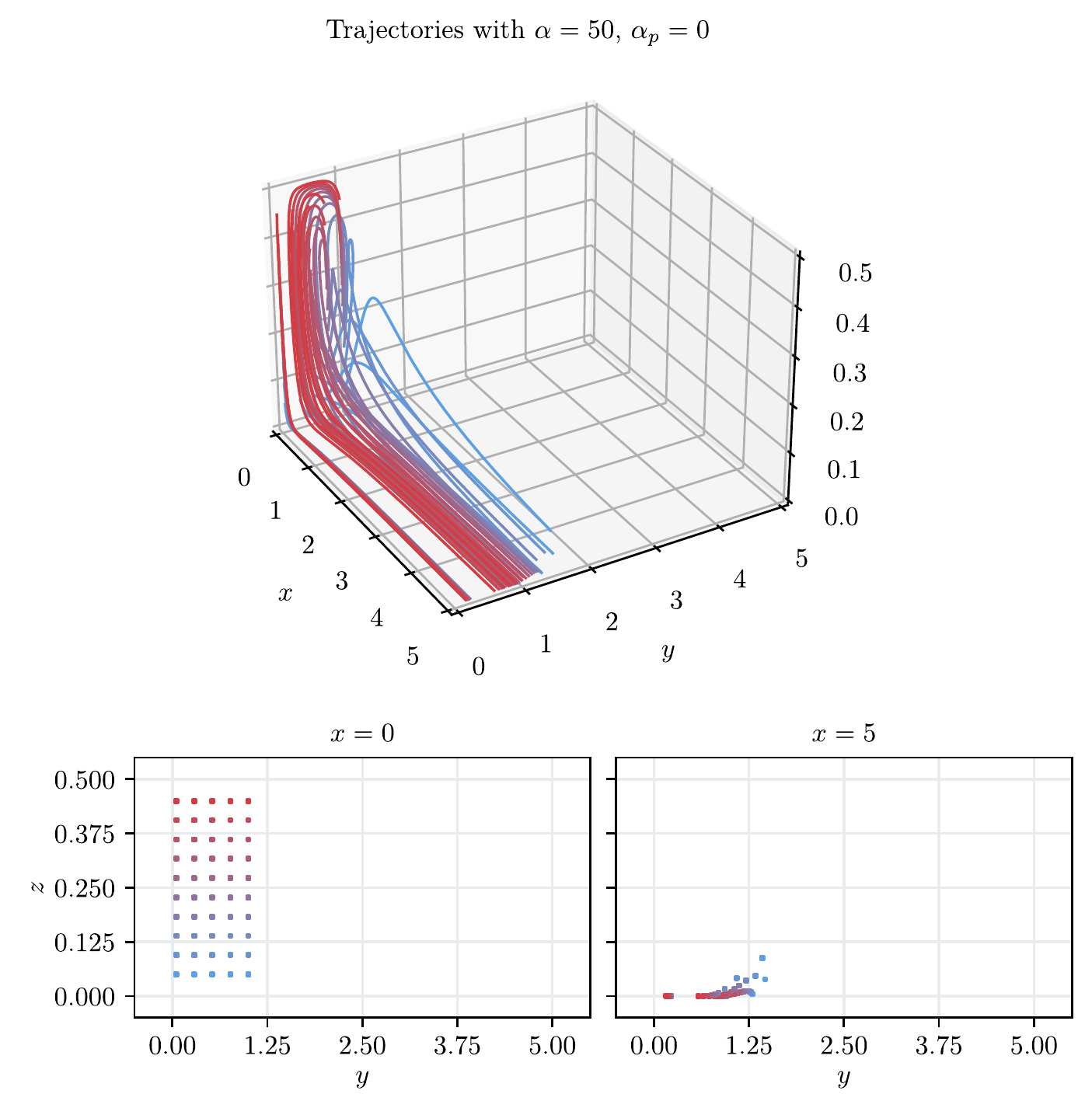}
    \caption{Example particle trajectories with large $\alpha$. Here, $\Delta=10$ and the solute concentration is lowest in the center of the channel. The model is valid for small $\alpha$, but we use a large value for visualization. Particle paths are colored by initial $z$-position.}\label{fig:trajectoryextreme}
\end{figure}

We show an extreme case, where $\alpha=50$, in \FIGTEXT{trajectoryextreme} to better demonstrate the convection rolls. Once again, though the theory is valid for small $\alpha$, we choose a large value to demonstrate the effect. The particles trace the velocity profile and demonstrate the existence of a vortex, where particles are drawn toward $y=0$ along the wall by the slip flow and outward from $y=0$ by the recirculating flow. The strength of the convection roll decays as $x$ increases because the solute diffuses and the concentration gradient diminishes.

\section{Details of experiments\label{sec:app:experimental-materials}}
The materials used in experiments are described in \TABTEXT{experimental-materials}.
{
\renewcommand{\arraystretch}{1.5}
\begin{table}[ht]
\centering
\begin{tabular}{l l}
Component & Origin\\\hline
\gls{ps} particles, $\DIM{a}=\SI{0.1}{\micro\meter}$ & Thermo Scientific{\texttrademark}, red fluorescent (catalog number R200)\\
\gls{cps} particles, $\DIM{a}=\SI{0.5}{\micro\meter}$ & Invitrogen FluoSpheres{\texttrademark}, red fluorescent\\
\gls{cps} particles,* $\DIM{a}=\SI{0.1}{\micro\meter}$ & Bangs Laboratories, green fluorescent (catalog number FCDG003)\\
\gls{aps} particles, $\DIM{a}=\SI{0.5}{\micro\meter}$ & Invitrogen FluoSpheres{\texttrademark}, yellow-green fluorescent\\
\gls{pdms} & SYLGARD{\texttrademark} 184 silicone elastomer kit\\
Glass & VWR{\texttrademark} microscope slides (catalog number 48300--026)
\end{tabular}%
\caption{Components used in experiments and their origins. *Used in zeta potentiometry experiments by Viet Sang Doan, University at Buffalo.}\label{tab:experimental-materials}
\end{table}
}

\section{Models for zeta potential and mobility\label{sec:app:zetamobilitymodels}}
\begin{figure}
    \centering
    \includegraphics{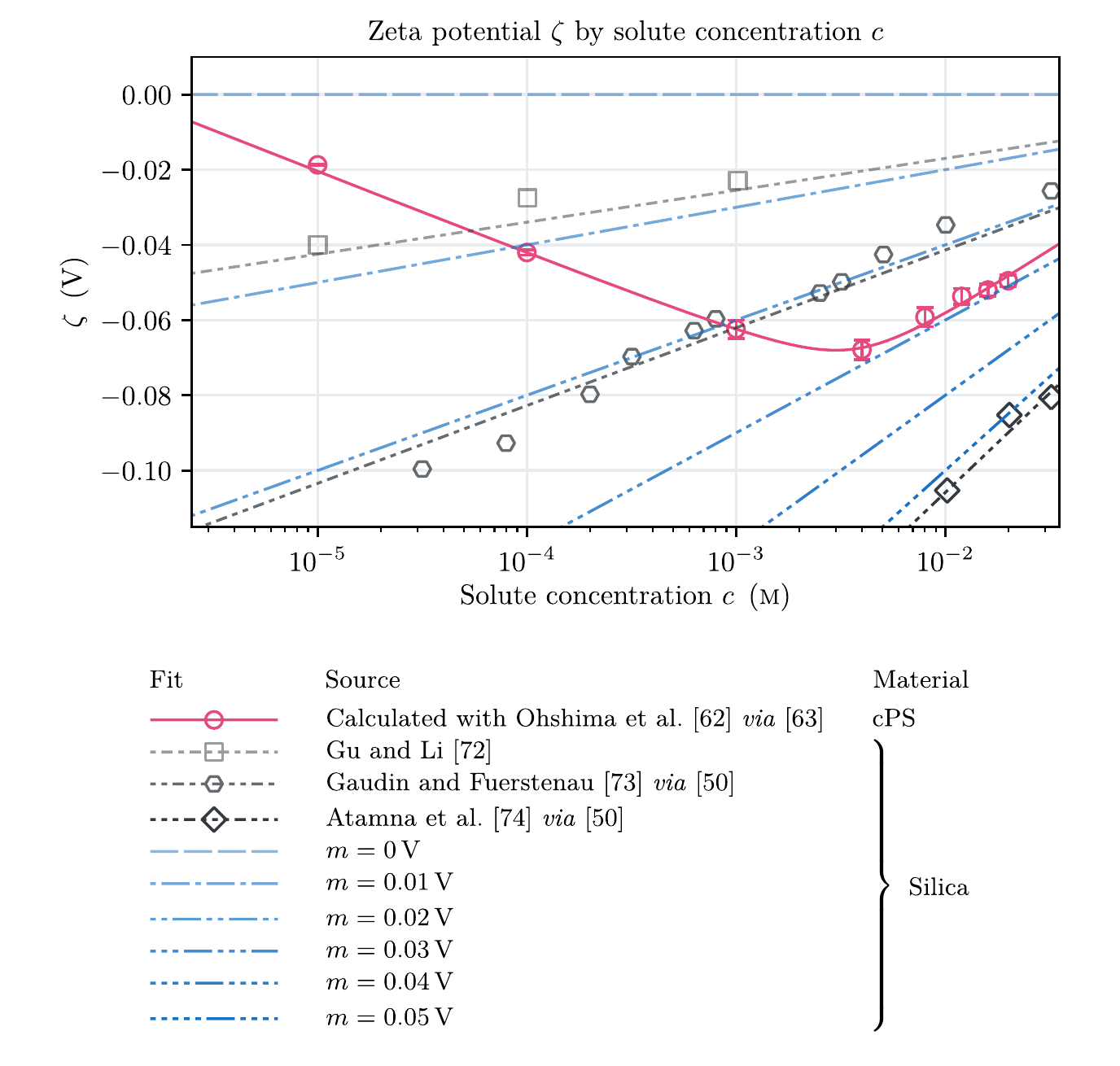}
    \caption{Zeta potential as a function of solute concentration. The particle zeta potential (pink) was calculated using electrophoretic mobility measurements obtained by Viet Sang Doan, University at Buffalo, and is for \SI{196}{\nano\meter} \gls*{cps} particles. Error bars indicate the standard deviation. We fit a line passing through the origin to experimental results for silica (note that the data shown on the plot may be truncated), as in the work of \citet{Kirby2004}. We fit a hyperbola to the zeta potential of the particles.}\label{fig:zeta-concentration}
\end{figure}

We implement a model with variable zeta potential, diffusiophoretic mobility, and diffusioosmotic mobility. First, we use electrophoretic mobility measurements of \gls{cps} particles (Viet Sang Doan, University at Buffalo) to calculate the particle zeta potential with the model of \citet{Ohshima1983}, as given by \citet{KirbyBook}. The result is shown in \FIGTEXT{zeta-concentration}. Notably, the magnitude of the particle zeta potential diminishes at low solute concentration, which is behavior not observed in published measurements of the zeta potential of silica \cite{Gu2000,Gaudin1955,Atamna1991}, for which there is a monotonic increase with decreasing solute concentration (also shown in \FIGTEXT{zeta-concentration}). Within the range of concentrations considered in experiments, it appears that there are two regimes for the particle zeta potential: At high concentration, the behavior is similar to that of glass, and the zeta potential decreases in magnitude with increasing concentration. At low concentration, the zeta potential increases in magnitude with increasing concentration. The two regimes appear approximately linear in $\ln \NONDIM{c}$, and we fit a hyperbola to obtain a continuous, empirical zeta potential. We propose a fit
\begin{equation}\label{eq:zeta-fit}
\begin{aligned}
\DIM{\zeta} &\approx -\DIMCONST{\alpha_0} + \DIMCONST{\alpha_1} \ln{\NONDIM{c}} + \sqrt{\DIMCONST{\alpha_2} + \DIMCONST{\alpha_3} \ln{\NONDIM{c}} + \DIMCONST{\alpha_4} \left( \ln{\NONDIM{c}} \right)^2}\quad{\rm or}\\
\DIM{\zeta_{\rm b}} &\approx \DIMCONST{m} \log_{10} {\NONDIM{c}}
\end{aligned}
\end{equation}
to data for the zeta potentials over the range of solute concentrations used in experiments. Here, $\DIM{\zeta}$ and $\DIM{\zeta_{\rm b}}$ are the zeta potentials of the particles and the boundary, respectively. The constants $\DIMCONST{\alpha_i}$, which we determine numerically, are given in \TABTEXT{constants}. The conductivity of a solution with $\NONDIM{c} \approx 0$ cannot be measured by the Litesizer{\texttrademark}, which introduces uncertainty in the modeled zeta potential at low solute concentrations.

\begin{table}[ht]
\centering
\renewcommand{\arraystretch}{1.5}
\begin{tabular}{l l}
Symbol & Value\\\hline
$\DIMCONST{\alpha_0}$ & $\SI{5.94e-2}{\volt}$\\
$\DIMCONST{\alpha_1}$ & $\SI{3.21e-3}{\volt}$\\
$\DIMCONST{\alpha_2}$ & $\SI{5.26e-3}{\volt\squared}$\\
$\DIMCONST{\alpha_3}$ & $\SI{1.84e-3}{\volt\squared}$\\
$\DIMCONST{\alpha_4}$ & $\SI{1.63e-4}{\volt\squared}$
\end{tabular}%
\caption{Constants associated with the fit to the calculated zeta potential as a function of solute concentration, described by \EQTEXT{zeta-fit}.}\label{tab:constants}
\end{table}

\begin{figure}
    \centering
    \includegraphics{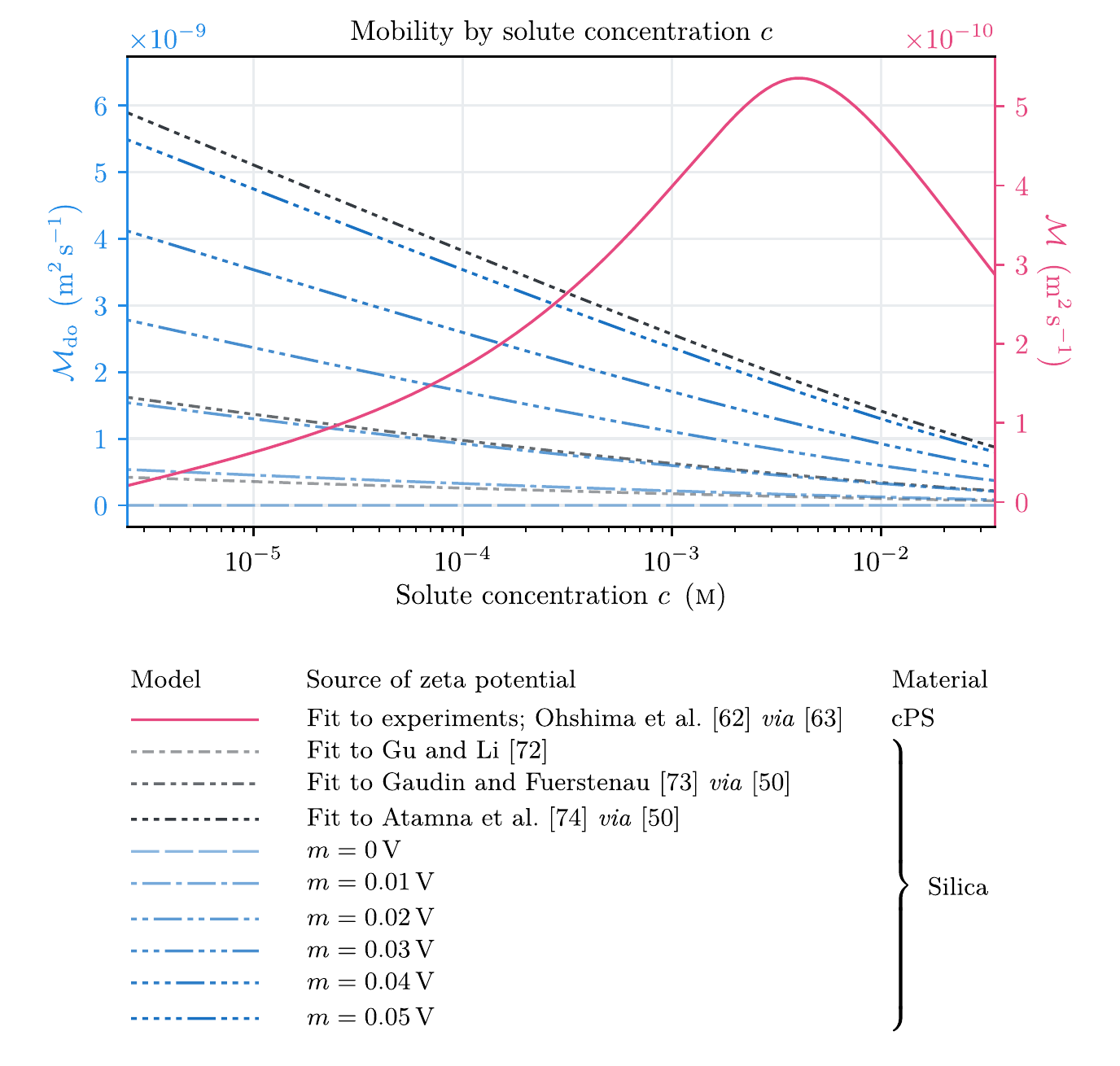}
    \caption{Diffusiophoretic and diffusioosmotic mobilities as a function of solute concentration. Note that the scale of the ordinate differs for each. The zeta potential of the wall is unbounded; this yields a significant diffusioosmotic mobility at low solute concentration. Consequently, we expect diffusioosmotic transport may be most significant at low solute concentration, which is consistent with our observations of convection rolls near the channel inlets and within $\left| \NONDIM{y} \right| < \frac{1}{6}$.}\label{fig:mobility-concentration}
\end{figure}

We now calculate the diffusiophoretic mobility $\DIM{{\mathcal{M}}}$ using the empirical fits to the zeta potential while accounting for finite-Debye length effects in diffusiophoresis. We use the model of \citet{Keh2000} for the mobility of the particles because it is valid for arbitrary values of the thickness ratio $\NONDIM{\lambda} = \left( \DIM{\kappa} \DIMCONST{a} \right)^{-1}$, where
\begin{equation}
\DIM{\kappa} = \sqrt{\frac{\DIMCONST{z}^2 \DIMCONST{e}^2 \DIMCONST{\sigma} \NONDIM{c}}{\DIMCONST{\varepsilon} \DIMCONST{{k_{\text{B}}}} \DIMCONST{T}}}
\end{equation}
is the inverse of the Debye length with valence $\DIMCONST{z}$, fundamental charge $\DIMCONST{e}$, and permittivity $\DIMCONST{\varepsilon}$; with a characteristic concentration of $\SI{1}{\Molar}$, the constant is $\DIMCONST{\sigma} \approx \SI{1.204e27}{\per\meter\cubed}$. Other models, such as that given by \citet{Prieve1984}, are valid only for $\NONDIM{\lambda} \ll 1$, which is violated at low solute concentrations, where the Debye length is comparable to the particle size. We model~\footnote{The form of the equation differs slightly from that given by \citet{Keh2000} to remain consistent with the conventions of recent works on diffusiophoresis and diffusioosmosis; in some papers, an additional factor of $4 \pi$ is present in the definition of permittivity, for which we use the convention $\DIMCONST{\varepsilon}=\NONDIMCONST{\varepsilon_{\rm r}} \DIMCONST{\varepsilon_0}$ with dielectric constant $\NONDIMCONST{\varepsilon_{\rm r}}$ and vacuum permittivity $\DIMCONST{\varepsilon_0}$ (\SEE{p.~5290}~of~\cite{Keh2000}).} the diffusiophoretic mobility as \cite{Keh2000}
\begin{equation}\label{eq:dp-mobility}
\DIM{{\mathcal{M}}} = \frac{\DIMCONST{\varepsilon}}{\DIMCONST{\mu}} \left[
\frac{\DIMCONST{{k_{\text{B}}}} \DIMCONST{T}}{\DIMCONST{z} \DIMCONST{e}} \NONDIM{\Theta_1} \NONDIMCONST{\beta} \DIM{\zeta} + \frac{1}{8} \NONDIM{\Theta_2} \DIM{\zeta}^2 + \ORDER{\DIM{\zeta}^3}
\right],
\end{equation}
where $\NONDIM{\beta} = \frac{\DIMCONST{{\mathcal{D}}_{+}} - \DIMCONST{{\mathcal{D}}_{-}}}{\DIMCONST{{\mathcal{D}}_{+}} + \DIMCONST{{\mathcal{D}}_{-}}}$ is the normalized diffusivity contrast, with the fits of \citet{Masliyah1994} to the functions $\NONDIM{\Theta_1}\left( \NONDIM{\lambda} \right)$ and $\NONDIM{\Theta_2}\left( \NONDIM{\lambda} \right)$:
\begin{equation}\label{eq:Thetas}
\begin{aligned}
\NONDIM{\Theta_1}\left( \NONDIM{\lambda} \right) &\approx 1 - \frac{1}{3} \left[ 1 + \num{0.07234} \left( \NONDIM{\lambda} \right)^{\num{-1.129}} \right]^{-1}\quad{\rm and}\\
\NONDIM{\Theta_2}\left( \NONDIM{\lambda} \right) &\approx 1 - \left[ 1 + \num{0.085} \left( \NONDIM{\lambda} \right)^{-1} + \num{0.02} \left( \NONDIM{\lambda} \right)^{-\num{0.1}} \right]^{-1}.
\end{aligned}
\end{equation}
The resulting mobilities are shown in \FIGTEXT{mobility-concentration}. The diffusiophoretic mobility appears to have a maximum; at higher or lower solute concentrations, it is diminished. In effect, models with constant diffusiophoretic mobility tend to overpredict particle migration at low or high concentrations and underpredict near the maximum mobility. Fits to the functions $\NONDIM{\Theta_{i}}$ perform well over a wide range of thickness ratios $\NONDIM{\lambda}$ and do not introduce significant error to the model. In the case of diffusioosmotic mobility, the surface is flat, so the radius $\DIM{a}$ is infinite and the thickness ratio $\NONDIM{\lambda}$ vanishes. Consequently, both $\NONDIM{\Theta_1}=1$ and $\NONDIM{\Theta_2}=1$ and the correction for finite Debye length is no longer relevant. Instead, we model the diffusioosmotic mobility as \cite{Shin2015}
\begin{equation}\label{eq:domobility}
\DIM{{\mathcal{M}}_{{{\text{do}}}}} = \frac{\DIMCONST{\varepsilon}}{\DIMCONST{\mu}}
\left( \frac{\DIMCONST{{k_{\text{B}}}} \DIMCONST{T}}{\DIMCONST{z} \DIMCONST{e}} \right)^2
\left\{
\frac{\DIMCONST{z} \DIMCONST{e}}{\DIMCONST{{k_{\text{B}}}} \DIMCONST{T}} \NONDIMCONST{\beta} \DIM{\zeta_{\rm b}}
+ 4 \ln \left[ \cosh{\left( \frac{\DIMCONST{z} \DIMCONST{e}}{4 \DIMCONST{{k_{\text{B}}}} \DIMCONST{T}} \DIM{\zeta_{\rm b}} \right)} \right]
\right\}.
\end{equation}
This expression is valid where the Debye layer is thin relative to the channel dimensions \cite{Shim2022}, which contributes to uncertainty in the dynamics of the system as $\NONDIM{c} \rightarrow 0$. We use a model for the boundary zeta potential where $\DIM{\zeta_{\rm b}} = \DIMCONST{m} \log_{10} \NONDIM{c}$, which is consistent with reported zeta potentials of silica \cite{Kirby2004a}, as shown in \FIGTEXT{zeta-concentration}, and vary the constant $\DIMCONST{m}$ because of uncertainty about the diffusioosmotic mobility at the glass surface. The maximum value of $\DIMCONST{m}$ associated with published results for the zeta potential--solute concentration relationship of silica, as shown in \FIGTEXT{zeta-concentration}, is approximately $\SI{0.05}{\volt}$; this informs the range we use in simulations.

\begin{figure}
    \centering
    \includegraphics[width=\textwidth]{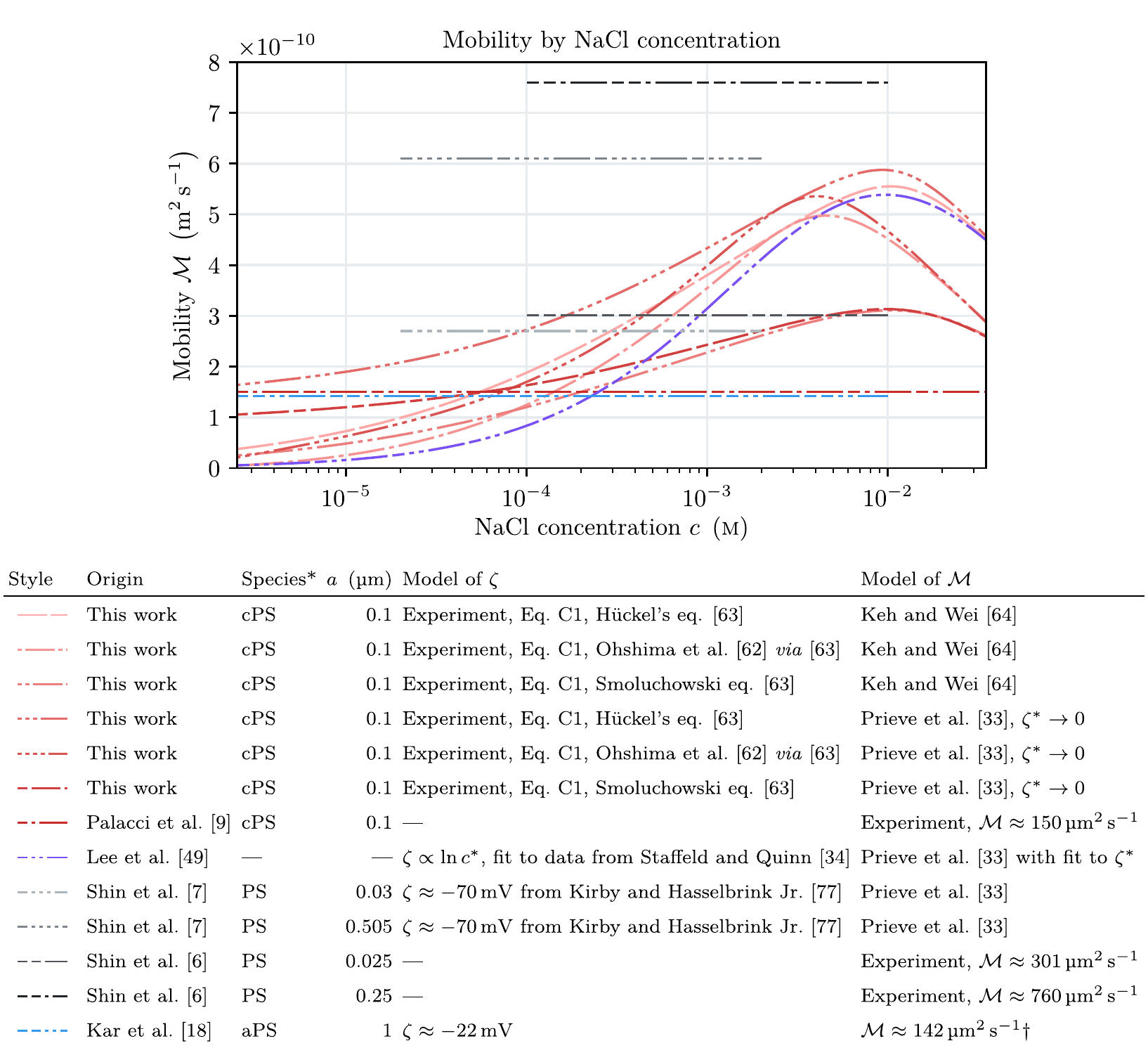}
    \caption{Various (nonexhaustive) estimates for diffusiophoretic mobility of particles in a solution of \ce{NaCl} at concentration $\DIM{c}$. Ultimately, the mobility takes a similar form, reaching a maximum at a particular solute concentration and decaying at higher or lower concentrations. *Some details of particle species, such as fluorescent coatings, are neglected. {\dag}Calculated with the expression given by the authors, which is analogous to \EQTEXT{domobility}.}\label{fig:mobility-comparison}
\end{figure}

We compare our model for diffusiophoretic mobility to several that have been used in previous works. Examples of models for the diffusiophoretic mobility of \gls{ps} particle species in \ce{NaCl} gradients are shown in \FIGTEXT{mobility-comparison}. Several authors consider a constant diffusiophoretic mobility, which is shown as a horizontal line with endpoints defined by the lowest and highest solute concentrations considered in experiments. Though the particles differ in size and surface chemistry, there is little variation in the values of $\DIM{{\mathcal{M}}}$---attributable, in part, to differences in thickness ratio $\NONDIM{\lambda}$---and the models for concentration-dependent mobility demonstrate similar behavior and have maxima at similar concentrations.

\section{Details of simulations}\label{sec:app:simdetails}
\begin{figure}[ht]
    \centering
    \includegraphics{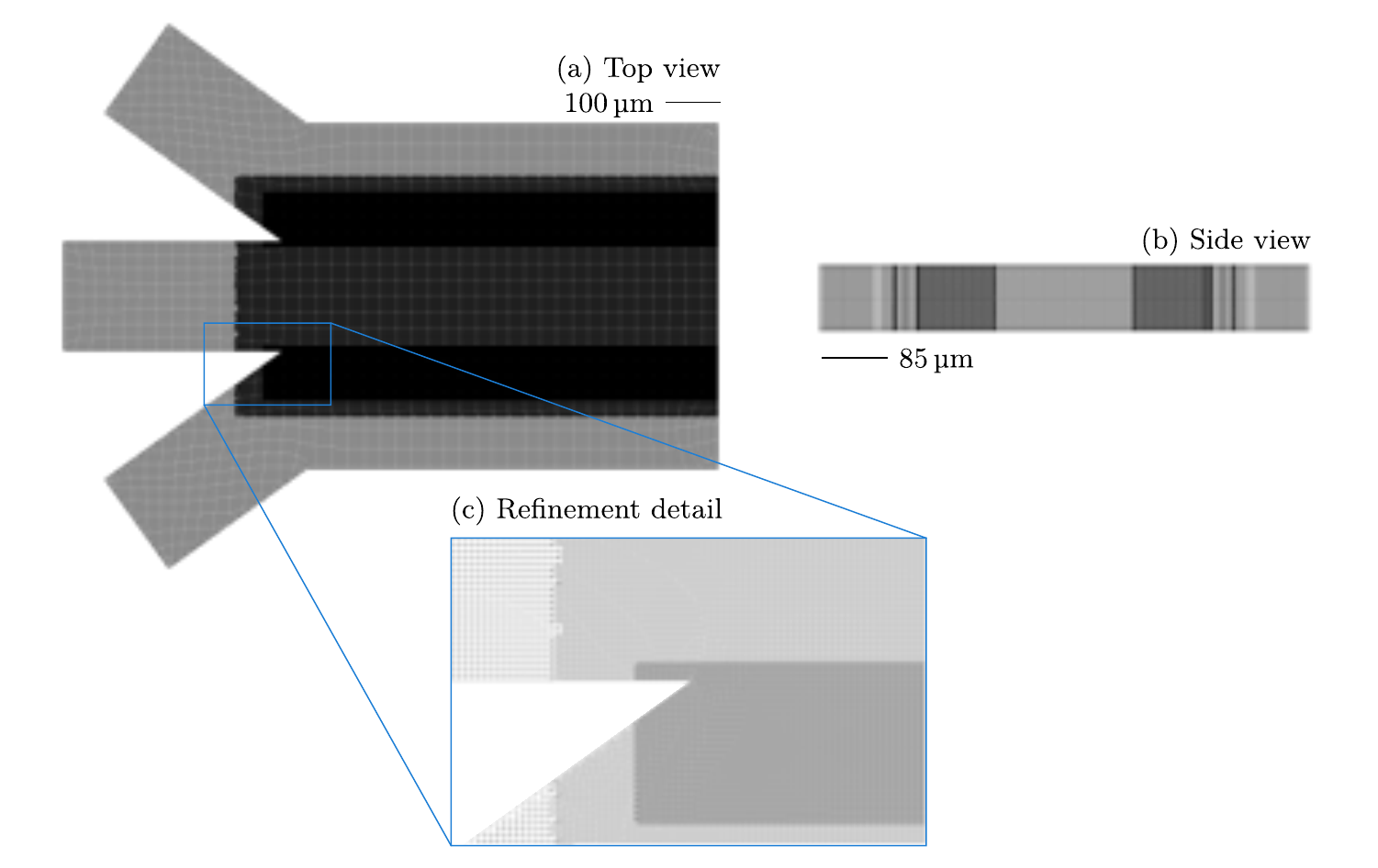}
    \caption{The mesh used for simulations, shown from the top in (a) and the side in (b). We have manually refined regions with large gradients in solute and particle concentrations, shown in (c).}\label{fig:mesh}
\end{figure}

The mesh we use for simulations is shown in \FIGTEXT{mesh}. We manually refined the mesh in regions where the solute concentration gradient is significant, such as the interfaces between solute streams. We do not simulate the full channel because it is computationally prohibitive; we simulate to $\DIM{x} \approx \SI{800}{\micro\meter}$, which is about $\SI{4}{\percent}$ of the channel length. We impose zero-gradient conditions for the solute and particle concentrations and the velocity at the outlet. We also fix the pressure at the outlet as $\NONDIM{p}=0$ and set the inlet velocity to be consistent with the flow rate of the syringe pumps. At the walls of the channel, we use zero-gradient conditions on the solute and particle concentrations.

\end{document}